\documentclass{aa}  
\usepackage{graphicx}
\usepackage{txfonts}
\usepackage{lipsum}
\usepackage{subcaption}
\usepackage{lscape}
\usepackage{placeins}
\usepackage{algorithm}
\usepackage{algpseudocode}
\usepackage[colorlinks=true, linkcolor=blue, citecolor=blue, urlcolor=black]{hyperref}

\begin{document}

   \title{RPCA: a resolved PSF reconstruction method for undersampled images}

   \author{Lin Nie
          \inst{1}
          \and
          Huanyuan Shan\inst{1}
          \thanks{hyshan@shao.ac.cn}
          \and
          Guoliang Li\inst{2}
	\and
	Tobías I. Liaudat\inst{3}
          \and 
          Chengliang Wei \inst{2}
	\and
	Peng Wei \inst{4}
          \and
          Tianmeng Zhang \inst{4}
          \and
          Hekun Li \inst{1}
          \and
          Lei Wang \inst{2}
          \and
          Han Qu \inst{2}
          \and
          Zhaojun Yan \inst{1}
          \and
          Nan Li \inst{4}
          \and
          You Wu \inst{4}
          \and
          Peng Jia \inst{5}
          }

   \institute{
        Shanghai Astronomical Observatory (SHAO), Nandan Road 80, Shanghai 200030, People's Republic of China
         \and
             Purple Mountain Observatory, Chinese Academy of Sciences, Nanjing, Jiangsu 210023, People's Republic of China
	\and
        IRFU, CEA, Université Paris-Saclay, F-91191 Gif-sur-Yvette, France
        \and
            National Astronomical Observatories, Chinese Academy of Sciences, Beijing 100101, People's Republic of China
        \and
            College of Physics and Optoelectronics, Taiyuan University of Technology, Taiyuan, 030024, China
             }

   \date{Received xx xx, xxxx; accepted xx xx, xxxx}

  \abstract
   {
   The Point-Spread Function (PSF) is a critical component in various astrophysical research areas, including galaxy morphology, photometric redshift determination, and weak lensing studies. As the next generation of space telescopes prepares for operations, including the China Space Station Telescope (CSST), accurate PSF modeling has become increasingly important. However, accurately modeling the PSF remains challenging, particularly for poorly sampled images.
   }
   {
   To construct reliable PSF models based on star images, we aim to develop a novel and robust method, which accounts for the pixel integration effect in modeling for reconstructing high-resolution PSFs from undersampled or poorly resolved images.
   } 
   {
{In this paper, we present Resolved Principal Components Analysis (RPCA), a novel approach that accounts for pixel integration effects to enable the direct reconstruction of resolved PSFs from undersampled images via an Expectation-Maximization PCA (EMPCA) technique.}
   }
   {
   Our method has been tested on a simulated PSF dataset for the China Space Station Telescope (CSST), where both observational and detector-induced effects were considered during simulation. The results demonstrate that RPCA achieves superior precision in PSF shape parameters compared to existing methods such as Multi-CCD (MCCD), Resolved Components Analysis (RCA), and PSFEx. Notably, RPCA {could reduce the bias of $\Delta R^2/R^2$ by approximately one order of magnitude with respect to MCCD, RCA and PSFEx.}
   }
   {
   RPCA exhibits robust stability under realistic observational conditions, including varying signal-to-noise ratios (SNRs) and contamination from cosmic rays and bad pixels. This suggests that RPCA could be a promising candidate for future CSST PSF modeling efforts.
}


   \maketitle

\nolinenumbers

\section{Introduction}\label{sec:intro}

The Point Spread Function (PSF) characterizes how point sources are imaged during observations, influenced by telescope optics, atmospheric turbulence (for ground-based facilities), and detector effects. The observed images of galaxies can be blurred and distorted by the PSF, making accurate PSF modeling a crucial step for astrophysical studies based on astronomical image analysis, such as galaxy morphology analysis \citep{2022ascl.soft01010R} and shape estimation in weak lensing \citep{2018ARA&A..56..393M,2024MNRAS.527.5206Y}. The precision of the PSF model is thus a key consideration in astronomical imaging studies. Several ambitious imaging surveys are planned or underway, including Euclid \citep{2012SPIE.8442E..0ZA,2025A&A...697A...1E}, LSST \citep{2019ApJ...873..111I}, CSST \citep{2019ApJ...883..203G}, and Roman \citep{10.1117/12.3020622}; therefore accurate PSF modeling would be one of the core tasks for these surveys.

Typically, the PSF model is reconstructed using star images (point sources) within the Field of View (FOV). The establishment of a PSF model primarily relies on sampled star images. However, this requirement becomes challenging to meet with undersampled images, where critical details are lost, making accurate modeling difficult \citep{2000PASP..112.1360A}. According to the Nyquist-Shannon Sampling Theorem, the theoretical form of the PSF cannot be accurately reconstructed from inadequately sampled star images. Poor sampling is particularly problematic for space-based telescopes like the Hubble Space Telescope (HST) and the James Webb Space Telescope (JWST), which usually have higher pixel resolution but smaller PSF sizes \citep{2000PASP..112.1360A}.

Various PSF modeling methods have been developed to address potential issues arising from poor sampling. Forward-modeling techniques, such as Tiny Tim for HST \citep{1995ASPC...77..349K} and WebbPSF for JWST and Roman \citep{2015ascl.soft04007P, 2017romanrept.1703L}, generate PSF models by simulating the imaging process. However, discrepancies between the model and data may arise due to variations in instrument configurations \citep{2018AAS...23115036H,2024AJ....167...58N}. An alternative approach involves reconstructing the PSF model directly from observational data (data-driven approaches). Despite being state-of-the-art, methods like PSFEx \citep{2011ASPC..442..435B}, Resolved Components Analysis (RCA) \citep{2016InvPr124001}, and Multi-CCD (MCCD) \citep{2021A&A...646A..27L} still exhibit PSF relative size errors exceeding 0.1 on images similar to those from Euclid \citep{2023InvPr..39c5008L,2023FrASS..1058213L}. Furthermore, complex structures in the PSF induced by diffraction effects are challenging to describe accurately with parametric models like the Moffat or Gaussian functions \citep{2001MNRAS.328..977T, 2015A&C....10..121R}. Recently, \citet{2023InvPr..39c5008L} introduced WaveDiff, a new PSF modeling framework that combines instrumental optical system information with observational data to achieve higher precision, although it does not yet account for detector effects such as electron diffusion.

Generally, accurate PSF model construction relies first on reliable PSF reconstruction at star positions, followed by interpolation to target positions. This paper {mainly focuses on the PSF reconstruction at star positions}. We introduce a novel method called Resolved Principal Components Analysis (RPCA) designed for undersampled images, decomposing the PSF model into a set of resolved orthonormal principal components. As a data-driven approach, RPCA constructs these principal components solely from star images, achieving super-resolution. We compare the precision of our reconstructed PSF model against state-of-the-art methods RCA and MCCD, as well as the widely used PSFEx, based on simulated CSST data. Section \ref{sec:methodology} describes the methodology of RPCA, while Section \ref{sec:test} demonstrates its performance relative to RCA, MCCD, and PSFEx. A summary is provided in Section \ref{sec:sums}.

\section{RPCA methodology} \label{sec:methodology}
Our method builds on subpixel shifting and downsampling, which are also employed by PSFEx, RCA, and MCCD.  Therefore, we first introduce the basic ideas of these methods in Section~\ref{sec:psfex}, and then introduce RPCA in Section~\ref{sec:RPCA}.
\subsection{PSFEx, RCA and MCCD}\label{sec:psfex}

To address the under-sampling problem, PSFEx employs the principle of super-resolution: unaliased components can, in principle, be recovered from images of multiple point sources randomly distributed with respect to the pixel grid \citep{1984ACVIP...1..317T}. Specifically, PSFEx assumes that the components follow a polynomial law:
\begin{equation}
\textrm{PSF}(x,y) = \sum_{d=1}^M \mathcal{S}_d \mathcal{A}_d(x,y),
\end{equation}
{where $(x, y)$ denotes the position on the charge-coupled device (CCD), $\mathcal{A}$ represents the position-dependent coefficients of the components $\mathcal{S}$, and $d$ indicates the polynomial order used to capture potential variations of the PSF across the field.  Here, M denotes the total number of monomials up to polynomial degree N.
For instance, if the maximum polynomial order N used in PSFEx is 2, then $M=6$ and $\mathcal{A} = \{c_0, c_1x, c_2y, c_3x^2, c_4xy, c_5y^2\}$, with $c_i$ typically set to 1 in PSFEx.} The components $\mathcal{S}$ are determined by minimizing the following cost function:
\begin{eqnarray}
\boldsymbol{L(\mathcal{A,S})} &=& \sum^{\rm nstar}_i \sum^{\rm npixel}_k \frac{\left[I_{ik} - f_i \Sigma_{d=1}^M \mathcal{A}_{id} \mathcal{F}(\mathcal{S}_{o,d} + \Delta \mathcal{S}_d)\right]^2}{\sigma^2_{ik}} \nonumber \\
&+& {\sum_{d=1}^M}(\tau \Delta \mathcal{S}_d)^2,
\end{eqnarray}
{where $I_{ik}$ is the value of the $k$-th pixel for the $i$-th star image}, $f_i$ is the flux estimated from $I_i$, $\mathcal{F}$ represents the subpixel shift and downsampling operation to align the model $\mathcal{S} = \mathcal{S}_{o,d} + \Delta \mathcal{S}_d$ with the data, $\mathcal{S}_{o,d}$ is the initial rough estimate derived from the point sources, and $\sigma$ is the pixel noise variance. In the second term, $\tau$ is a regularization scalar that ensures smoothness in $\Delta \mathcal{S}$.

A similar decomposition of the PSF is also employed in RCA. However, RCA incorporates additional regularizers into the PSF modeling process \citep{2016InvPr124001,2021A&A...646A..27L}: i) low-rankness by limiting the number of {PSF feature components}; ii) positivity of the reconstructed PSFs; iii) sparsity of the PSF representation in an appropriate basis; and iv) spatial constraints to enforce structural consistency within the $\mathcal{A}$ matrix. MCCD further extends this approach by introducing global and local components based on RCA to construct a unified model for multiple CCD images in a single exposure. Both RCA and MCCD have demonstrated superior PSF modeling performance compared to PSFEx in tests using simulated Euclid telescope PSFs \citep{2023InvPr..39c5008L}.

The spatial variation constraints and smooth regularization applied in these methods render the ill-posed problem caused by undersampling and noise solvable while effectively capturing the characteristics of PSF field variations. {There is another class of methods (such as principal component analysis (PCA) \citep{Pearson01111901}) that reconstruct the PSF by decomposing the observed stellar images into a set of PSF features without explicitly introducing spatial correlations. As a result, these methods need to adopt additional approaches to capture the spatial variations of the PSF. However, this also leaves room for combining such methods with other techniques in PSF spatial interpolation.} Several studies have investigated the performance of various interpolation methods on different observational datasets, highlighting that the issue of PSF interpolation must be carefully addressed \citep{2012MNRAS.419.2356B, 2012MNRAS.427.2572C, 2013A&A...549A...1G, 2017AJ....153..197L, 2020A&A...636A..78S, 2024ApJ...972..120A}. Taking these factors into account, we relax the spatial constraints in PSF reconstruction.

PCA, a powerful non-parametric data compression method, has been widely used in PSF reconstruction for weak lensing studies \citep{10.1098/rsta.2015.0202, 2016RAA....16..139L}. However, classical PCA lacks the ability to handle poorly resolved images. Nevertheless, the orthogonality property of the principal components (PCs) provides a potential avenue for addressing the ill-posed nature of PSF reconstruction in a super-resolution framework.
\subsection{RPCA}\label{sec:RPCA}
\subsubsection{Motivation of RPCA}

In the PCA framework, the PSF corresponding to the $i$-th star, denoted as $\textrm{PSF}_i$, can be decomposed into a set of orthogonal principal components (PCs):

\begin{equation}
\textrm{PSF}_i = \sum_{l=1}^{\rm npc} C_{il} P_l,
\end{equation}
where $C_{il}$ represents the coefficient of the principal component $P_l$ { and npc represents the number of PCs used}. The coefficients $C$ and the principal components $P$ can be determined using methods such as the Karhunen-Loève (K-L) transform \citep{2002SPIE.4836..350L} or singular value decomposition (SVD) \citep{10.1098/rsta.2015.0202}. Additionally, \citet{2012PASP..124.1015B} introduced an efficient alternative approach called Expectation-Maximization PCA (EMPCA), which is particularly well-suited for handling noisy and incomplete data$-$a common scenario in astronomical imaging. EMPCA has recently been applied to PSF reconstruction tasks with promising results \citep{2016RAA....16..139L,2021MNRAS.503.4436N,2024AJ....167...58N}.

In this work, we adopt EMPCA as the base algorithmic methodology for constructing PCs with super-resolution capabilities. Following the EMPCA procedure, we define a loss function as follows:

\begin{equation}
\label{EMPCA}
\boldsymbol{L(C,P)} = \sum_{i=1}^{\rm nstar} \sum_{k=1}^{\rm npixel} \frac{\left[I_{ik} - \sum_{l=1}^{\rm npc} C_{il} \mathcal{F}(P_l)\right]^2}{\sigma_{ik}^2},
\end{equation}
where $\mathcal{F}$ denotes the subpixel shifting and downsampling operation, which uses a Lanczos function to interpolate the value at the downsampled position by using the oversampled pixel value \citep{2011ASPC..442..435B}, applied to the oversampled principal component $P_l$. The key challenge now lies in selecting an appropriate $\mathcal{F}$ and solving for $P$ and $C$. 

\subsubsection{Linear operation of subpixel shifting and down-sampling}
Due to the random sampling of the centers of observed and undersampled star images on the pixel grid, it is necessary to first align the center of $P$ to the stars' positions in the super-sampled grid, followed by down-sampling the principal component to match the resolution of the observed data.

The subpixel shifting operation is also known as image interpolation. The value $y^{\prime}$ at the interpolated position $(u, v)$ can be expressed as a weighted summation of the original image $y$:
\begin{equation}
\label{conv}
y^{\prime}(u, v) = \sum_{u^{\prime}, v^{\prime}} K(u - u^{\prime}, v - v^{\prime}) y(u^{\prime}, v^{\prime}),
\end{equation}
where $K$ is the interpolation kernel, and $(u^{\prime}, v^{\prime})$ and $(u, v)$ represent the coordinates before and after interpolation, respectively. Equation~\ref{conv} can also be interpreted as a discrete convolution operation between $K$ and $y$. If we treat a 2D (two-dimensional) image as a 1D (one-dimensional) vector, the interpolated image $\mathbf{I^{\prime}}$ can be represented as the matrix product of an interpolation matrix $\mathbf{S}$ and the original image vector $\mathbf{I}^{\ast}$: 
\begin{equation}
\mathbf{I^{\prime}} = \mathbf{S} \mathbf{I}^{\ast}.
\end{equation}
The matrix $\mathbf{S}$ can be directly constructed based on Equation~\ref{conv} (see Section \ref{sec:appa} for details). A similar process for handling interpolation has been applied in \citet{2021MNRAS.501.1282J}. In this paper, the interpolation kernel used is derived from the convolution kernel between the shifted and unshifted images using Fourier Transform \citep{1999SPIE.3661..728C, 2021MNRAS.503.4436N}.

Next, let us consider the down-sampling operation. Analogously, the down-sampled image $\mathbf{I}$ can be expressed as the matrix product of a down-sampling matrix $\mathbf{D}$ and the super-sampled image vector $\mathbf{I^{\prime}}$\citep{2015A&A...575A..86N}. For example, if the oversampling factor is 2, the value of $\mathbf{I}$ at position $(0, 0)$ in the down-sampled image is the summation of the four pixels $[(0, 0), (0, 1), (1, 0), (1, 1)]$ in the super-sampled image. In this case, the weights for these four pixels are set to 1, while the weights for all other pixels are 0. Consequently, the observed pixelized image $\mathbf{I}$ and the pseudo super-sampled image $\mathbf{I^{\ast}}$ can be related as:
\begin{equation}
\mathbf{I} = \mathbf{D} \mathbf{S} \mathbf{I^{\ast}},
\end{equation}
where $\mathbf{D}$ and $\mathbf{S}$ represent the down-sampling and interpolation operation matrices  (see Section \ref{sec:appa} for details), respectively. As a result, Equation~\ref{EMPCA} can now be rewritten as:
\begin{equation}
\label{RPCA0}
\boldsymbol{L(C,P)}  = \sum_{i=1}^{\rm nstar} \left[\mathbf{I}_i - \sum_{l=1}^{\rm npc} C_{il} \mathbf{D} \mathbf{S}_i \mathbf{P}_l \right]^T \mathbf{V}_i^{-1} \left[\mathbf{I}_i - \sum_{l=1}^{\rm npc} C_{il} \mathbf{D} \mathbf{S}_i \mathbf{P}_l \right],
\end{equation}

where $\mathbf{D}$, $\mathbf{S}_i$, and $\mathbf{P}_l$ denote the down-sampling matrix, the sub-pixel shifting matrix, and the $l$-th principal component vector, respectively, and `$T$' denotes the transpose of matrix. Here, $C_{il}$ represents the coefficient for the $l$-th component of the $i$-th star. $\mathbf{V}_i$ is a diagonal matrix representing the pixel variances of the corresponding star image vector $\mathbf{I}_i$, assuming independent noise between pixels. {A key refinement in Equation \ref{RPCA0} compared to the PSFEx formalism is the explicit inclusion of the pixel integration process through the down-sampling matrix $\mathbf{D}$.}

{Furthermore, any masked pixels (e.g., those affected by cosmic rays or hot pixels) are assigned a weight of zero in $\mathbf{V}_i^{-1}$. Since astronomical noise typically follows a Poisson-Gaussian distribution, we employ a Gaussian approximation with flux-dependent weights ($\sigma_{i,j}^2 = \sigma_{\text{bg}}^2 + \sigma_{\text{poisson},i,j}^2$, where $\sigma_{\text{bg}}$ and $\sigma_{\text{poisson},i,j}$ represent the background noise and Poisson noise, respectively; see e.g., \citealt{2021MNRAS.503.4436N}) to account for photon noise. This heteroscedastic approach is statistically robust for high-count star samples and offers significant computational advantages over a full Poisson likelihood.}

\subsubsection{Smooth constraints}

As demonstrated in \citet{2012PASP..124.1015B} and \citet{2021MNRAS.503.4436N}, the orthonormality of principal components (PCs) is achieved manually after solving the PCs by minimizing the $\chi^2$ equation. Consequently, additional constraints must be imposed on the PCs to ensure that the solution exists and yields reasonable results. 

Following a similar approach to PSFEx, we incorporate the regularization term $\lambda_1 |\mathbf{P}|^2$ into Equation~\ref{RPCA0} {in order to avoid potentially unstable solutions, {where `||' denotes the summation of the squared values of the matrix inside}. To further improve the numerical stability of the solution and suppress overfitting, we assume that the PSF profile is continuous. Therefore, we require both the gradient and the second derivative of $\mathbf{P}$ to be smooth.} To enforce this assumption, we introduce two additional regularization terms: $\lambda_2 |\nabla \mathbf{P}|^2$ and $\lambda_3 |\Delta \mathbf{P}|^2$, where ``$\nabla$'' and ``$\Delta$'' represent the gradient and second derivative operators, respectively. With these regularization terms included, we could have:

\begin{eqnarray}
\label{RPCA1}
\boldsymbol{L(C,P)} &=& \sum_{i=1}^{\rm nstar} \left[\mathbf{I}_{i} - \sum_{l=1}^{\rm npc} C_{il} \mathbf{D} \mathbf{S}_i \mathbf{P}_{l}\right]^T \mathbf{V}_{i}^{-1} \left[\mathbf{I}_{i} - \sum_{l=1}^{\rm npc} C_{il} \mathbf{D} \mathbf{S}_i \mathbf{P}_{l}\right] \nonumber \\
&+& \lambda_1 |\mathbf{P}|^2 + \lambda_2 |\nabla \mathbf{P}|^2 + \lambda_3 |\Delta \mathbf{P}|^2  \nonumber 
\\
&=& \sum_{i=1}^{\rm nstar} \left[\mathbf{I}_{i} - \mathbf{D} \mathbf{S}_i \mathbf{P} \mathbf{C}_i\right]^T \mathbf{V}_{i}^{-1} \left[\mathbf{I}_{i} - \mathbf{D} \mathbf{S}_i \mathbf{P} \mathbf{C}_i\right] \nonumber \\
&+& \lambda_1 |\mathbf{P}|^2 + \lambda_2 |\nabla \mathbf{P}|^2 + \lambda_3 |\Delta \mathbf{P}|^2,
\end{eqnarray}
where the weighted summation of $\mathbf{P}$ and $\mathbf{C}$ is reorganized into linear combinations in the second equation \citep{2021MNRAS.503.4436N}. In practical applications, the operators ``$\nabla$'' and ``$\Delta$'' can be expressed as first-order and second-order difference matrices, denoted as $\boldsymbol{M}_{\text{grad}}$ and $\boldsymbol{M}_{\text{Lap}}$, {which are the gradient and Laplacian matrices \citep{2002dip..book.....G,2006aida.book.....S}, respectively.}  ]

{The gradient and Laplacian operators at each pixel can be expressed as 2-D matrices and subsequently flattened into vectors, following the same procedure described in Appendix~\ref{sec:appa}. Accordingly, the first- and second-order finite differences at a given pixel can be written as inner products between the corresponding flattened operator vectors and $\mathbf{P}$.
By stacking the operator vectors from all pixels, we construct two 2-D matrices, $\boldsymbol{M}_{\text{grad}}$ and $\boldsymbol{M}_{\text{Lap}}$. The regularization terms $|\nabla \mathbf{P}|^2$ and $|\Delta \mathbf{P}|^2$ can then be written in quadratic form as: 
$(\boldsymbol{M}_{\text{grad}} \mathbf{P})^T \boldsymbol{M}_{\text{grad}}\mathbf{P}$, $(\boldsymbol{M}_{\text{Lap}}\mathbf{P} )^T \boldsymbol{M}_{\text{Lap}}\mathbf{P} $ respectively.}

{Owing to undersampling effects and noise contamination, imposing reasonable physical constraints is helpful for obtaining more stable and accurate solutions. An important consideration for CSST-like PSFs is that the PSF intensity decreases from the center toward the periphery. Therefore, additional smoothness constraints are introduced to $\boldsymbol{M}_{\text{grad}}$ and $\boldsymbol{M}_{\text{Lap}}$:
\begin{eqnarray}
\label{RPCA2}
\boldsymbol{L(C,P)} &=& \sum_{i=1}^{\rm nstar} \left[\mathbf{I}_{i} - \mathbf{D} \mathbf{S}_i \mathbf{P} \mathbf{C}_i\right]^T \mathbf{V}_{i}^{-1} \left[\mathbf{I}_{i} - \mathbf{D} \mathbf{S}_i \mathbf{P} \mathbf{C}_i\right] \nonumber \\
&+& \lambda_1 \mathbf{P}^T\mathbf{P} + {\lambda}_2 (\boldsymbol{M}^\mathcal{S}_{\text{grad}} \mathbf{P})^T \boldsymbol{M}^\mathcal{S}_{\text{grad}}\mathbf{P} \nonumber \\
&+&  {\lambda}_3 (\boldsymbol{M}^\mathcal{S}_{\text{Lap}}\mathbf{P} )^T \boldsymbol{M}^\mathcal{S}_{\text{Lap}}\mathbf{P} ,
\end{eqnarray}
where $\boldsymbol{M}^\mathcal{S}_{\text{grad}}$ and $\boldsymbol{M}^\mathcal{S}_{\text{Lap}}$, in which each row corresponding to an individual pixel is multiplied by the spatial weight respectively, represent the gradient and second derivative regularization terms with spatial weighting. In this work, this weighting function is proportional to $r^2$ (where $r$ is the pixel distance from the PSF center), effectively enforcing stronger smoothness constraints in the outer regions where the signal is weaker\footnote{The CSST optical system features an off-axis design with a three-mirror anastigmat (TMA) telescope, which eliminates diffraction spikes from supporting structures. This justifies our assumption of a profile that is steep at the core and increasingly smooth toward the periphery.}. While this specific radial weighting is optimized for the CSST PSF, the functional form of $\mathcal{S}$ can be adapted for other instruments. 
Furthermore, the regularization parameters $\lambda_1$, $\lambda_2$, and $\lambda_3$ play a role analogous to $1/\sigma_{\phi}^{2}$ in Eq.~1.6 of \citet{2011ASPC..442..435B}, where $\sigma_{\phi}\sim10^{-2}$ was found to provide a good compromise by effectively controlling noisy cases without detectable smoothing of PSFs in high signal-to-noise data. We therefore set $\lambda_1=\lambda_2=\lambda_3=10^{4}$ to ensure numerical stability.

\subsection{Solving $\mathbf{C}$ and $\mathbf{P}$}

To solve the principal components (PCs) $\mathbf{P}$ and their corresponding coefficients $\mathbf{C}$, we adopt the same solving procedures as described in \citet{2021MNRAS.503.4436N}.

Minimizing Equation~\ref{RPCA2} with respect to $\mathbf{C}_i$ for the $i$-th star reduces to solving the following system of equations:
\begin{eqnarray}
\label{solveC}
\frac{\partial \boldsymbol{L}}{\partial \mathbf{C}_i} & = & 2(\mathbf{D}\mathbf{S}_i \mathbf{P})^T \mathbf{V}_{i}^{-1} (\mathbf{I}_{i} - \mathbf{D}\mathbf{S}_i \mathbf{P} \mathbf{C}_i) = 0, \nonumber \\
(\mathbf{D}\mathbf{S}_i \mathbf{P})^T \mathbf{V}_{i}^{-1} \mathbf{I}_{i} & = & (\mathbf{D}\mathbf{S}_i \mathbf{P})^T \mathbf{V}_{i}^{-1} (\mathbf{D}\mathbf{S}_i \mathbf{P}) \mathbf{C}_i, \nonumber \\
\mathbf{C}_i & = & \left[ (\mathbf{D}\mathbf{S}_i \mathbf{P})^T \mathbf{V}_{i}^{-1} (\mathbf{D}\mathbf{S}_i \mathbf{P}) \right]^{-1} (\mathbf{D}\mathbf{S}_i \mathbf{P})^T \mathbf{V}_{i}^{-1} \mathbf{I}_{i}.
\end{eqnarray}

Equation~\ref{solveC} is a linear system in $\mathbf{C}$ and $\mathbf{P}$ that allows the coefficients $\mathbf{C}_i$ for each star to be solved directly from the constructed matrices $\mathbf{I}$ and $\mathbf{P}$.
And in this work, the PCs are solved one by one to maximize the variance explained by each PC before proceeding to the next \citep{2021MNRAS.503.4436N}, rather than solving all PCs simultaneously. 

Similarly, to solve for the $m$-th order PC $\mathbf{P}_m$, we take the partial derivative of {the loss function} with respect to $\mathbf{P}_m$:
\begin{eqnarray}
\label{solveP}
\frac{\partial \boldsymbol{L}}{\partial \mathbf{P}_m} &=& 2[\sum_{i=1}^{\rm nstar} -C_{im} (\mathbf{D}\mathbf{S}_i)^T \mathbf{V}_i^{-1} (\mathbf{I}_i - C_{im} \mathbf{D}\mathbf{S}_i \mathbf{P}_m) \nonumber \\
&+& \lambda_1 \mathbf{P}_m 
+ {\lambda}_2 (\boldsymbol{M}^\mathcal{S}_{\text{grad}})^{T} \boldsymbol{M}^\mathcal{S}_{\text{grad}} \mathbf{P}_m \nonumber \\
&+& \lambda_3 (\boldsymbol{M}^\mathcal{S}_{\text{Lap}})^{T} \boldsymbol{M}^\mathcal{S}_{\text{Lap}} \mathbf{P}_m] \nonumber \\
&=& 0, \nonumber \\ 
\sum_{i=1}^{\rm nstar}C_{im} (\mathbf{D}\mathbf{S}_i)^T \mathbf{V}_i^{-1} \mathbf{I}_i &=& [ \sum_{i=1}^{\rm nstar}C_{im}^2 (\mathbf{D}\mathbf{S}_i)^T \mathbf{V}_i^{-1} (\mathbf{D}\mathbf{S}_i) \nonumber \\
&+& \lambda_1 \mathcal{I} + \lambda_2 (\boldsymbol{M}^\mathcal{S}_{\text{grad}})^{T} \boldsymbol{M}^\mathcal{S}_{\text{grad}} \nonumber \\
&+& \lambda_3 (\boldsymbol{M}^\mathcal{S}_{\text{Lap}})^{T} \boldsymbol{M}^\mathcal{S}_{\text{Lap}} ] \nonumber \\
&\times& \mathbf{P}_m, \nonumber  \\
\mathbf{P}_m &=& [ \sum_{i=1}^{\rm nstar}C_{im}^2 (\mathbf{D}\mathbf{S}_i)^T \mathbf{V}_i^{-1} (\mathbf{D}\mathbf{S}_i) \nonumber \\
&+& \lambda_1 \mathcal{I} + \lambda_2 (\boldsymbol{M}^\mathcal{S}_{\text{grad}})^{T} \boldsymbol{M}^\mathcal{S}_{\text{grad}} \nonumber \\
&+& \lambda_3 (\boldsymbol{M}^\mathcal{S}_{\text{Lap}})^{T} \boldsymbol{M}^\mathcal{S}_{\text{Lap}} ]^{-1} \nonumber \\
&\times& [\sum_{i=1}^{\rm nstar}C_{im} (\mathbf{D}\mathbf{S}_i)^T \mathbf{V}_i^{-1} \mathbf{I}_i],
\end{eqnarray}
where  $\mathcal{I}$ represents the identity matrix. {Equation~\ref{solveP} is also a linear system in $\mathbf{C}$ and $\mathbf{P}$. Therefore, $\mathbf{C}$ and $\mathbf{P}$ can be solved iteratively using Equation~\ref{solveC} and Equation~\ref{solveP}.} To obtain all PCs, we start by solving the first component $\mathbf{P}_1$, and subsequent components are solved based on the residual image data after removing variations explained by previously solved PCs: 
\begin{eqnarray}
\label{updateI}
\mathbf{I}_i = \mathbf{I}_i - \sum_{l=1}^{\rm N_{\text{solved}}} C_{il} \mathbf{D}\mathbf{S}_i \mathbf{P}_l.
\end{eqnarray}

After all PCs are solved using Equation~\ref{solveP}, they must be orthogonalized manually because noise in the data may break the orthogonality between PCs {and the amplitude of $\rm \boldsymbol{P}$ and $\rm \boldsymbol{C}$ are degenerate}. In this work, we use the Gram-Schmidt orthonormalization \citep{GSorth} procedure to orthogonalize the PCs. Finally, both $\mathbf{P}$ and $\mathbf{C}$ are solved iteratively, and the solving process is implemented using PyTorch in this work. Algorithm~\ref{alg:rpca} shows the pseudocode for solving $\mathbf{P}$ and $\mathbf{C}$. {Based on a trade-off between computational cost and accuracy, we find that 20 iterations are sufficient, and therefore we use 20 iterations to compute $\mathbf{P}$ and $\mathbf{C}$ in this work.} 

Once $\mathbf{P}$ and $\mathbf{C}$ are calculated, the PSF models at other positions could be interpolated by using low order polynomials to fit the spatial variations of the coefficients $\boldsymbol{C}$ across the field:
\begin{equation}
\label{polyC}
C_{il}(x_i, y_j) = a_{00} + a_{10}x + a_{01}y + a_{20}x^2 + a_{11}xy + a_{02}y^2 + \cdots
\end{equation}
where $(x_i, y_i)$ are the coordinates of the $i$-th star, and $a_{\text{xx}}$ represents the coefficients of the polynomials. This approach is widely used in PSF interpolation \citep{2005ApJ...618...46J,2007PASP..119.1403J,2017AJ....153..197L}.  A detailed investigation of alternative interpolation methods will be left for future work.

\begin{algorithm}
\caption{$\mathbf{P}$ and $\mathbf{C}$ solving process}
\label{alg:rpca}
\begin{algorithmic}[1]
\Statex Iterations
\State{Processing(): Constructing $\boldsymbol{D}$ and $\boldsymbol{S}$ matrix.} 
\State{$\boldsymbol{P}$ and $\boldsymbol{C}$ initialization: random value or start from PCA\footnotemark}
\While{iteration time $< \; \text{max iteration}$ or $\mathbf{L} >$ threshold}
    \For{$k = 1$ to $\text{max iteration}$}
        \For{$m = 1$ to $\text{npc}$}
            \State {$\boldsymbol{P_{m(k+1)}} \gets \boldsymbol{C_{i(k)}}$ (equation \ref{solveP}) }
            \State{update data matrix (equation \ref{updateI})}
        \EndFor
        \State{$\boldsymbol{P}$ orthogonalization}
        \For{$i = 1$ to $\text{nstar}$}
            \State {$\boldsymbol{C_{i(k+1)}} \gets \boldsymbol{P_{(k+1)}}$ (equation \ref{solveC}) }
        \EndFor
    \EndFor
\EndWhile
\State \Return $\boldsymbol{P}$ and $\boldsymbol{C}$
\end{algorithmic}
\end{algorithm}
\footnotetext{The initial values of $\boldsymbol{P}$ and $\boldsymbol{C}$ in this work are calculated from PCA.}

\section{PSF reconstruction test} \label{sec:test}

The China Space Station Telescope (CSST) is a planned 2-meter space telescope that will survey approximately 17,500 deg$^2$ of the sky in the wavelength range of {[255$\sim$1000]} nm across seven filter bands over a span of about 10 years. With a field of view of approximately 1.1 deg$^2$, CSST is designed to achieve an image resolution of $0.074^{\prime\prime}$ per pixel and a PSF full width at half maximum (FWHM) of $0.13^{\prime\prime}$\footnote{\url{https://nadc.china-vo.org/csst-bp/article/20230707113736}}. The PSF size relative to the pixel scale is comparable to that of HST\footnote{\url{https://hst-docs.stsci.edu/acsihb/chapter-5-imaging/5-6-acs-point-spread-functions}}%
\textsuperscript{,}%
\footnote{\url{https://hst-docs.stsci.edu/wfc3ihb/chapter-6-uvis-imaging-with-wfc3/6-6-uvis-optical-performance}}, implying that CSST may also encounter undersampling issues. In this section, we test our method using simulated PSF data for CSST.

\subsection{Test on high-fidelity simulated CSST PSF data}
\label{sec:PSF}

In this section, we use a high-fidelity dataset of simulated CSST PSFs {for the Main Survey Camera (MSC)} to test the performance of our method. The fiducial PSF data includes various optical aberrations such as mirror surface roughness, fabrication errors, CCD assembly errors, gravitational distortions, and thermal distortions, among others. Considering the wavelength range {[255$\sim$1000 nm]} across multiple bands and the stellar density expected for CSST, we randomly selected 100 stars, which approximates the number of PSF stars per CCD in real observations \citep{2018MNRAS.481.1149Z}. These stars are chosen from the `$u$' ({322$-$396 nm}) and `$z$' ({846$-$1000 nm}) bands to evaluate our method's capability in handling undersampled images.

\subsubsection{Data description}
\label{sec:cstdata}
We construct our test samples from the simulated oversampled PSFs, where one oversampled pixel corresponds to 0.25 pixels at the instrument resolution. We first perform sub-pixel scale random shifts on the oversampled PSFs (where one instrument resolution pixel corresponds to four pixels) using interpolation. The shifted PSFs are then downsampled to the target resolution. Next, random flux values are assigned to these samples. Finally, Poisson noise according to the flux in each pixel and background Gaussian noise are added to the PSF images. Six datasets with different Signal-to-Noise Ratios (SNRs) are generated to evaluate the performance of our method under various data qualities.

\begin{figure*}
\centering
\includegraphics[width=1\textwidth]{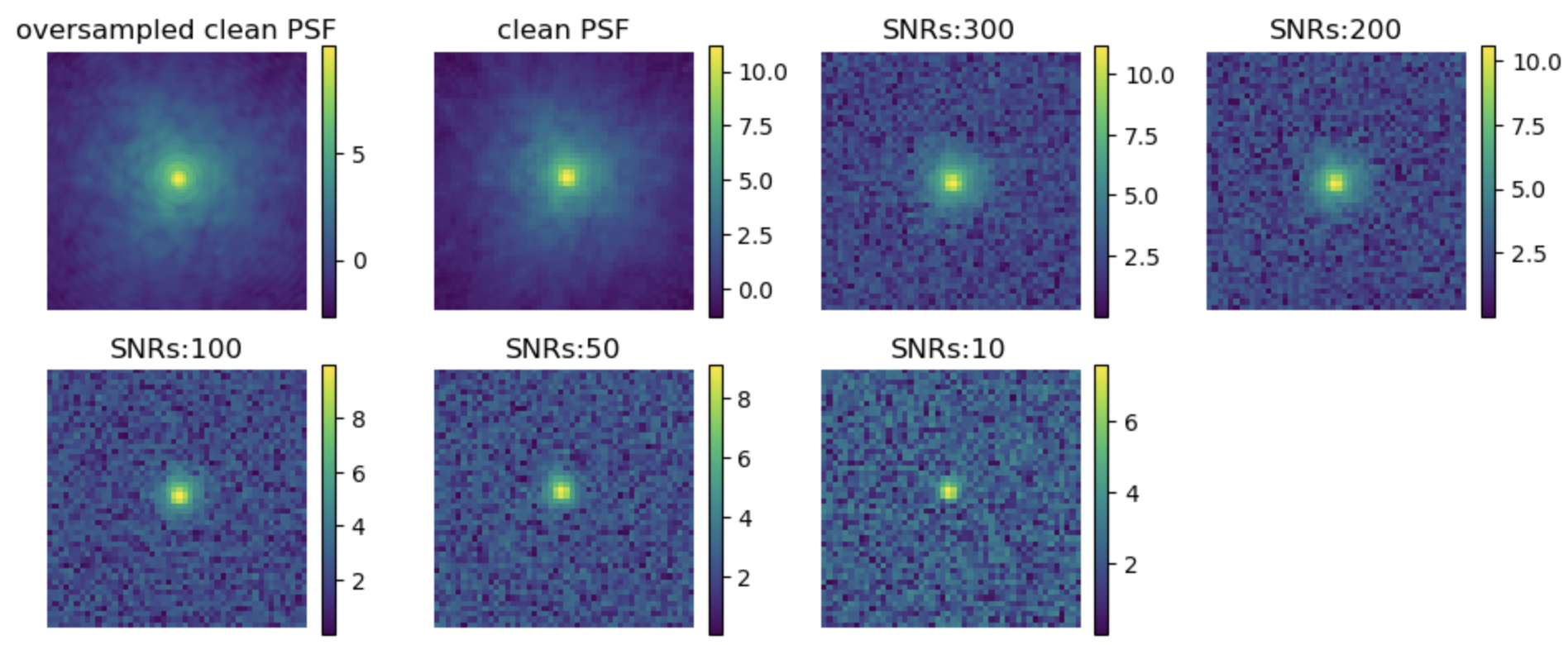}
\caption{Examples of star images for the oversampled PSF, and PSFs at instrument resolution with SNRs of 300, 200, 100, 50, and 10 in the `z' band. The pixel values have been rescaled for better visual performance.}
\label{star_sample}
\end{figure*}

\begin{figure*}
\centering
\includegraphics[width=1\textwidth]{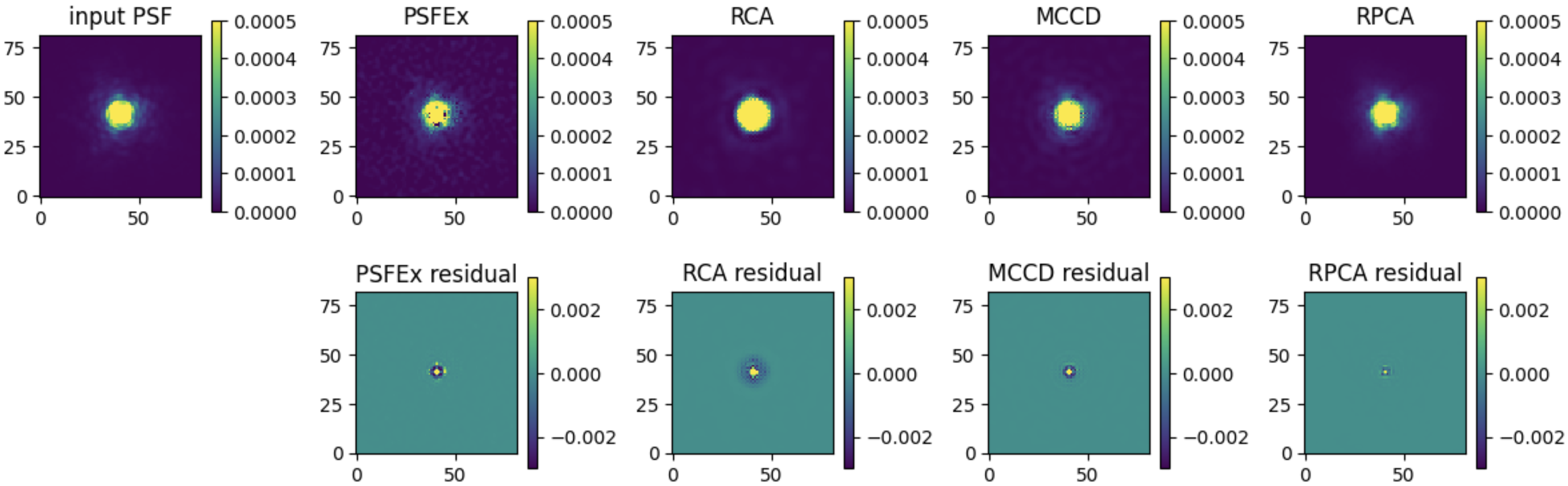}
\caption{Interpolated PSF models and corresponding residuals between the noise-free input PSF and the results from PSFEx, MCCD, RCA and RPCA, respectively, in the `z' band. The upper left panel shows the input PSF, while the other four panels display the interpolated PSFs from PSFEx, MCCD, RCA and RPCA, respectively. The corresponding residuals are shown in the lower panels. The pixel values have been rescaled for better visual performance.}
\label{residuals}
\end{figure*}

\begin{figure*}
\centering
\includegraphics[width=1\textwidth]{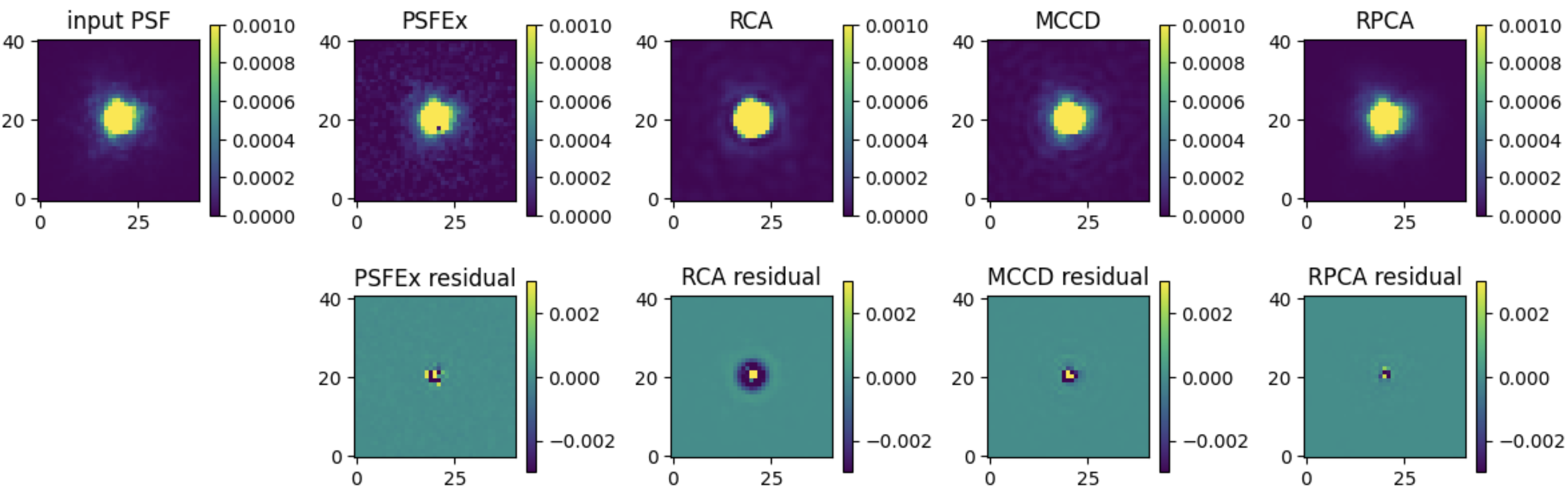}
\caption{The same as Fig.~\ref{residuals} but for the oversampled PSFs.}
\label{eresiduals}
\end{figure*}

Figure~\ref{star_sample} shows examples of star images with different SNRs in the `$z$' band. Meanwhile, the oversampled data are preserved for comparisons of the reconstructed PSFs with super-resolution. Additionally, RCA and MCCD, as state-of-the-art PSF reconstruction methods, are used as reference comparisons in this section. All test datasets are organized into three-dimensional arrays to match the input data format required by the three methods.

\subsubsection{PSF construction results}
\label{sec:comp}

In this work, we adopt an oversampling factor of 2 for the three test methods due to potential memory requirements and computational time constraints. Our method uses 20 principal components (PCs), which are generally sufficient for PSF reconstruction. {For PSF interpolation, we use third-order polynomials, which are expected to be robust for CSST-like PSF fields, to model the spatial variations of the coefficients $\boldsymbol{C}$ across the field.}

To determine appropriate parameter configurations for our dataset, we tested different parameter settings for both RCA and MCCD and selected the configurations that provided the best performance in this work. Finally, we adopt ${n\_comp\_loc = 10, ksig = 1}$ for RCA, and ${n\_comp\_loc = 5, d\_hyb\_loc = 3, ksig = 3}$ for MCCD on the simulated PSF dataset used in this section (see Section~\ref{sec:para-check} for details). It is worth noting that MCCD has the capability to reconstruct the PSF across the entire field of view (FOV) at once. However, since the wavelength range varies across different CCDs within the FOV for CSST, we perform PSF reconstruction on each CCD separately.  {Furthermore, to ensure consistency between our method and PSFEx (version 3.24.2), we manually mask all sources except the PSF stars in the input images for PSFEx. Additionally, we set the parameters ``-PSFVAR$\_$DEGREES 3 -PSF$\_$PIXELSIZE 0.5 -PSF$\_$SAMPLING 0.5'' in PSFEx to reconstruct oversampled PSFs, solving for 10 components, and we find that PSFEx performance is not sensitive to the ``-PSFVAR$\_$DEGREES'' for higher orders on our dataset. Finally, we use the interpolation module in \citet{2021A&A...646A..27L} to predict the PSF at interpolated positions.}

Next, we {interpolate} the PSF models {at $\sim$ 800 non-training positions, where the true PSFs are extracted from the simulation directly} using RPCA, MCCD, RCA and PSFEx based on the noisy datasets with different Signal-to-Noise Ratios (SNRs). Figure \ref{residuals} shows an example of the interpolated PSFs from RPCA, RCA, MCCD {and PSFEx} {at one of the interpolated positions}, along with their corresponding residuals based on the SNR=100 dataset. From the upper panels of Figure~\ref{residuals}, it is evident that MCCD and RPCA exhibit better consistency in the outer regions of the PSF compared to {RCA and PSFEx}, which show unexpected structures in the PSF. The lower panels of Figure~\ref{residuals} reveal that RPCA achieves the lowest residuals, and the residual structures from MCCD, RCA and PSFEx are more prominent than those from RPCA.

Additionally, Figure~\ref{eresiduals} presents an example of the {interpolated} oversampled PSFs from RPCA, {PSFEx, RCA and MCCD}, along with their corresponding residuals based on the SNR=100 dataset. The diffraction structures at the periphery of the oversampled PSF are {also} reconstructed by RPCA, whereas these features are not clearly visible in the results from {RCA and PSFEx}. The residuals show similar trends to those observed in Figure~\ref{residuals}.

\begin{figure*}
\includegraphics[width=0.5\textwidth]{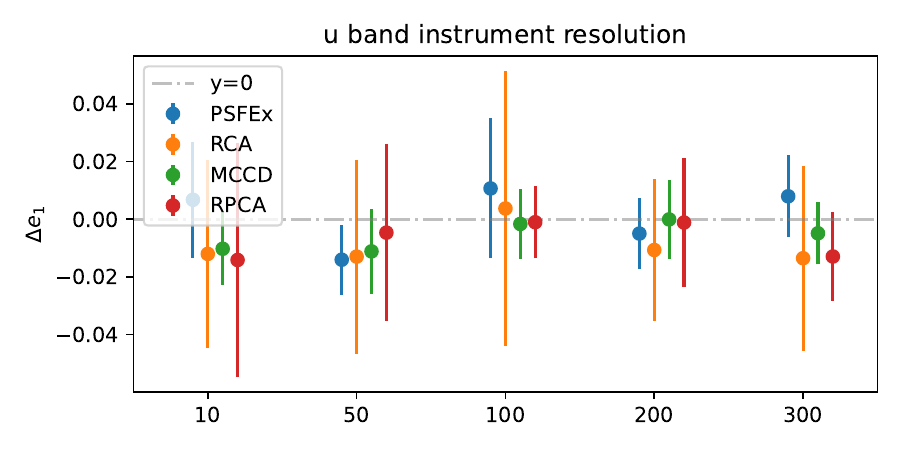}
\includegraphics[width=0.5\textwidth]{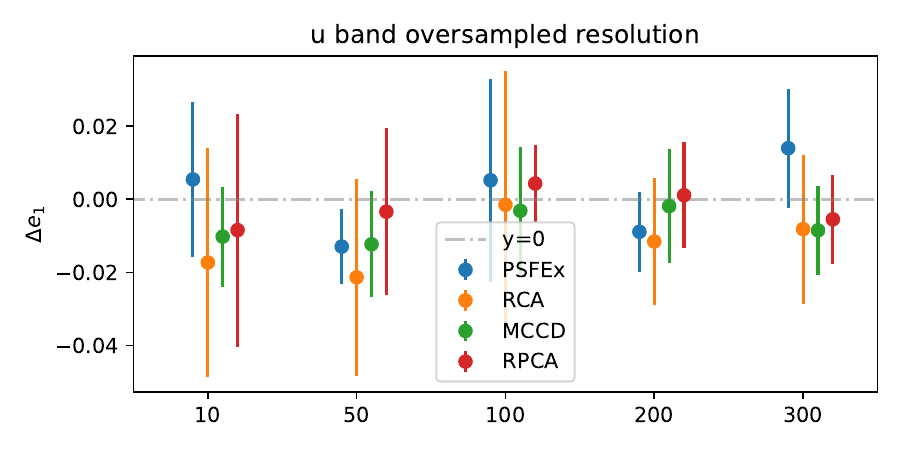}
\includegraphics[width=0.5\textwidth]{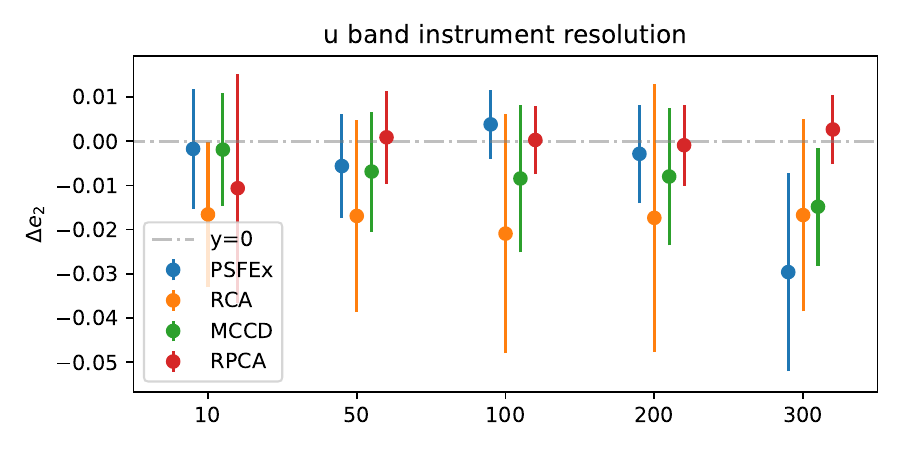}
\includegraphics[width=0.5\textwidth]{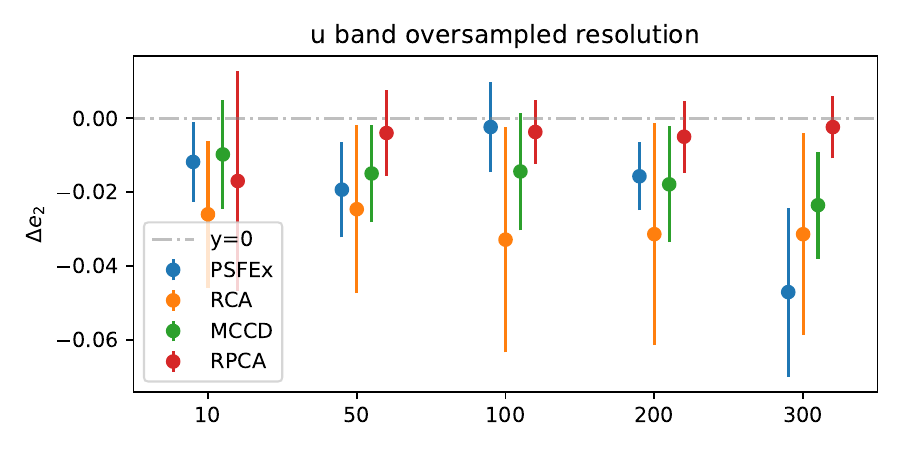}
\includegraphics[width=0.5\textwidth]{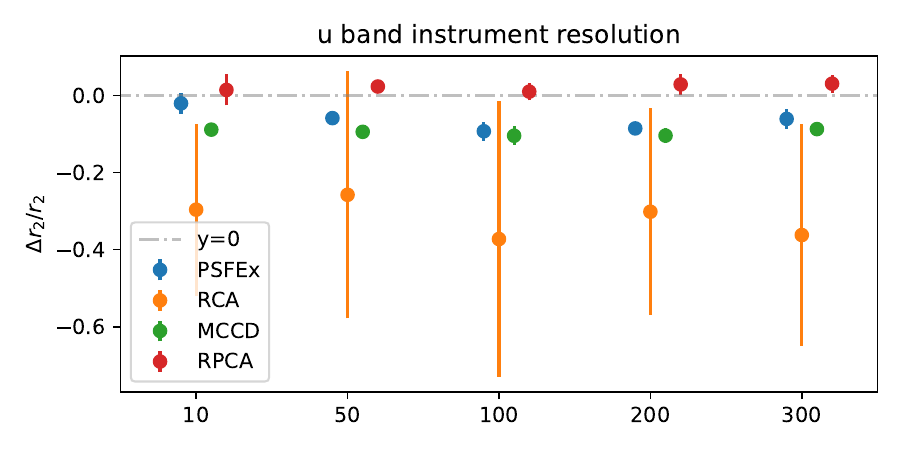}
\includegraphics[width=0.5\textwidth]{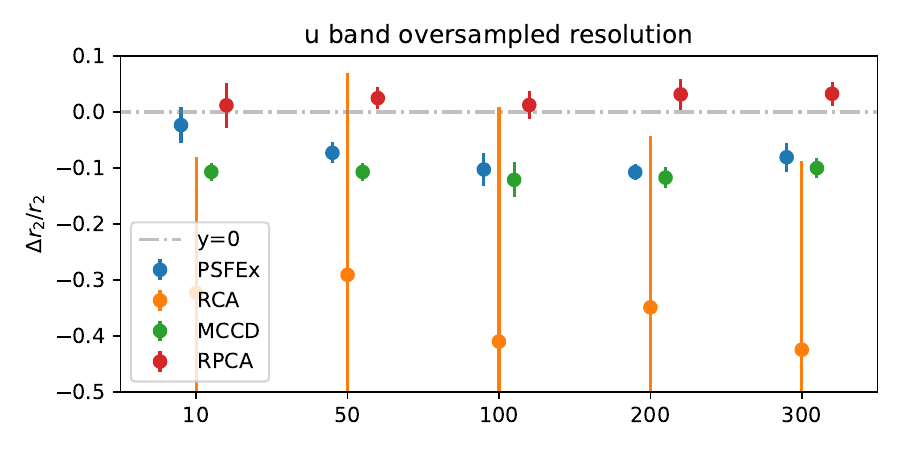}
\caption{Statistical shape residuals of  PSFEx (blue), RCA (orange), MCCD (green), and RPCA (red) are shown for datasets with SNRs of 10, 50, 100, 200, 300, and random values between 100 and 300. The panels from top to bottom represent the shape residuals: $\Delta e_1$, $\Delta e_2$, and $\Delta R^2/R^2$, respectively. The left and right panels show the residuals calculated from the undersampled and oversampled data, respectively. The error bars indicate the $1\sigma$ width of the distributions.}
\label{ushapes}
\end{figure*}

\begin{figure*}
\sidecaption
\parbox[b]{120mm}{\includegraphics[width=120mm]{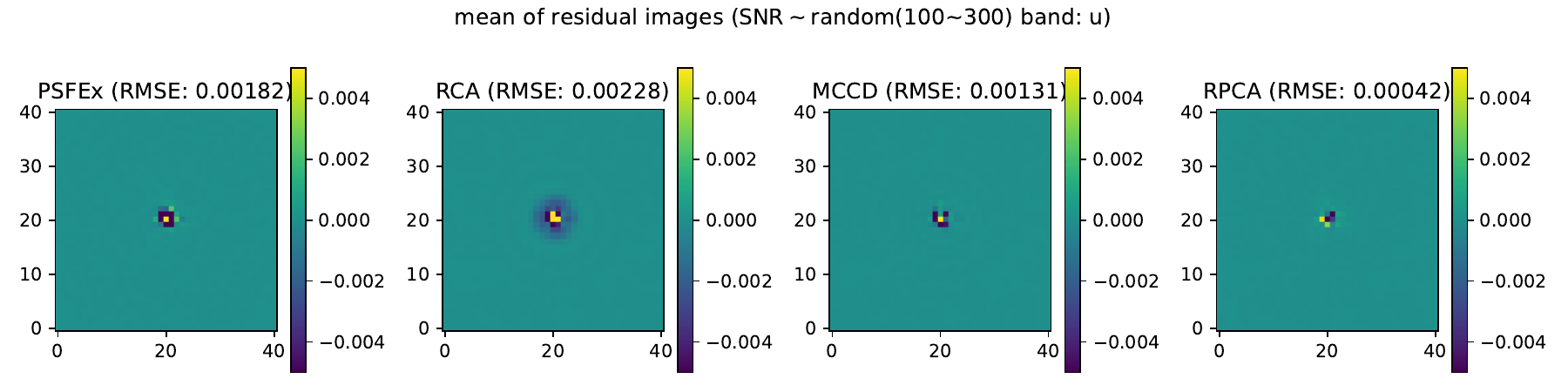}\\[2mm]
\includegraphics[width=120mm]{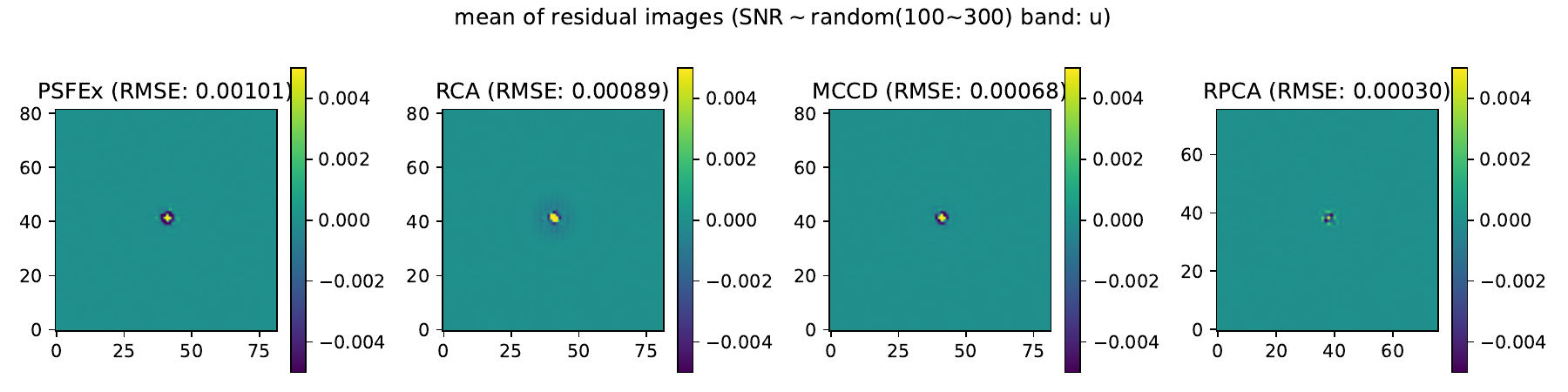}}
\caption{Mean residual images of the four methods for SNRs randomly distributed between 100 and 300 in the 'u' band. Upper panels: residuals of instrument resolution. Lower panels: residuals of oversampled resolution.}
\label{uRMSE}
\end{figure*}

\begin{figure*}
\sidecaption
\parbox[b]{120mm}{\includegraphics[width=120mm]{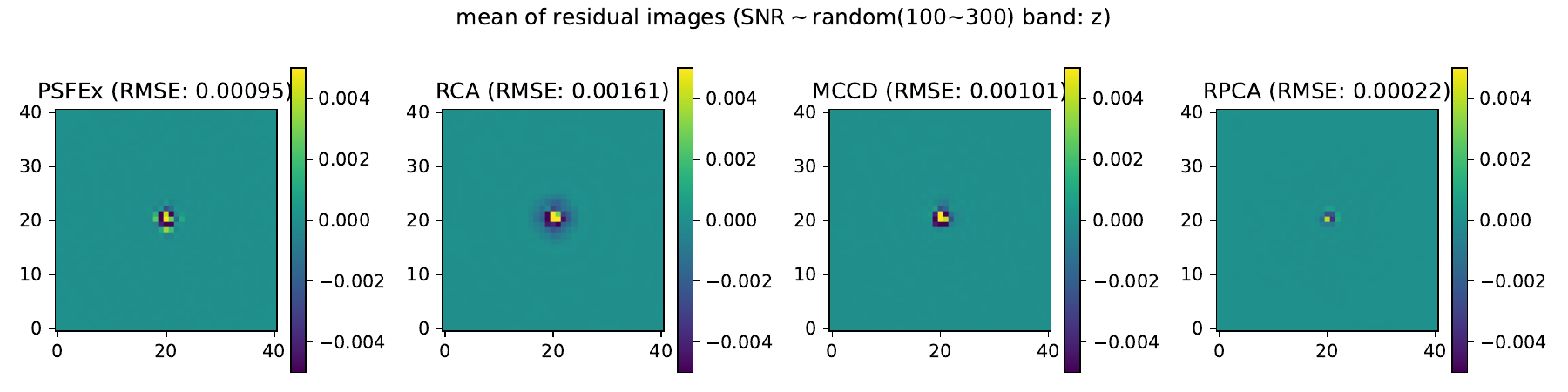}\\[2mm]
\includegraphics[width=120mm]{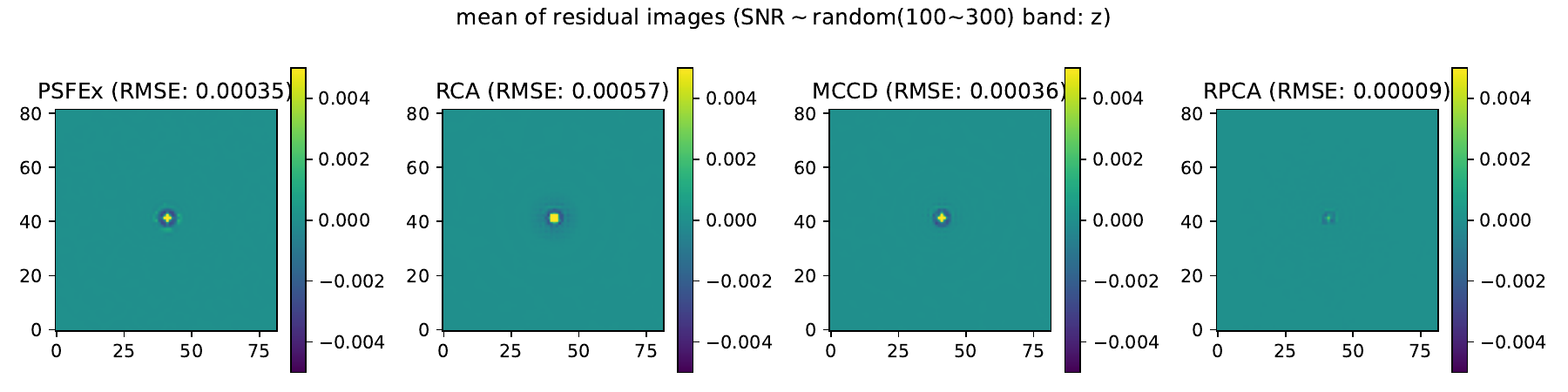}}
\caption{Mean residual images of the four methods for SNRs randomly distributed between 100 and 300 in the 'z' band. Upper panels: residuals of instrument resolution. Lower panels: residuals of oversampled resolution.}
\label{zRMSE}
\end{figure*}

\begin{figure}
\includegraphics[width=0.45\textwidth]{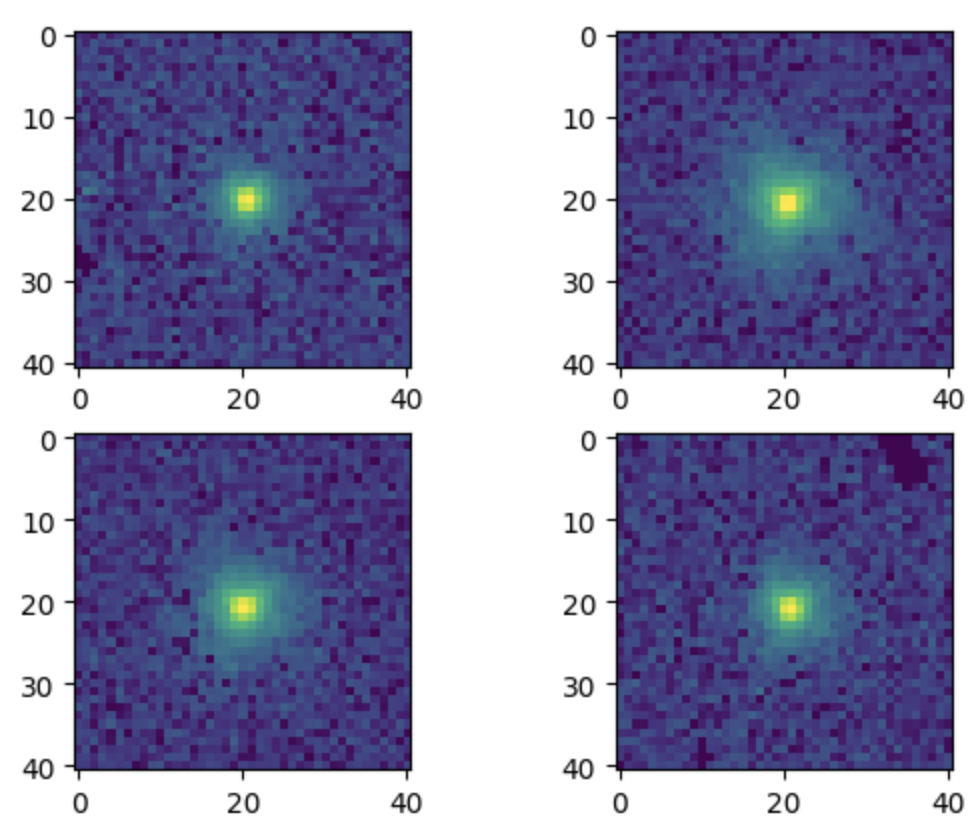}\\
\caption{Examples of extracted PSF star images from CSST MSC simulation data.}
\label{mscstars}
\end{figure}

\begin{figure}
\includegraphics[width=0.45\textwidth]{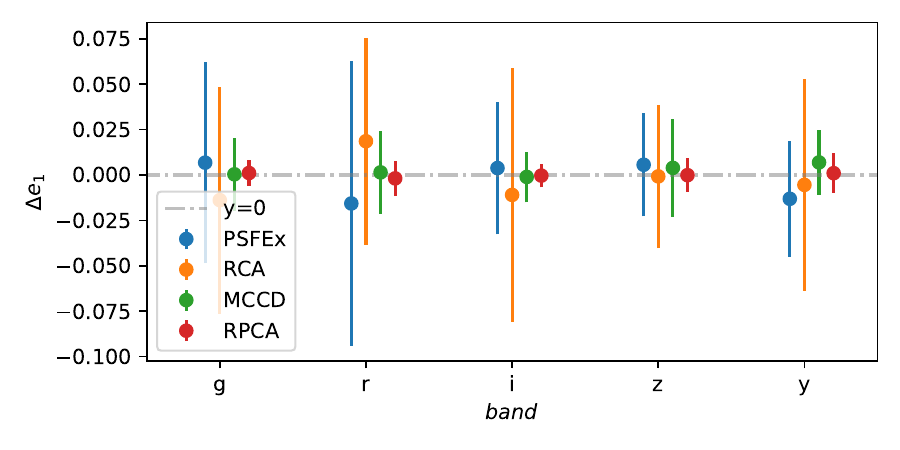}\\
\includegraphics[width=0.45\textwidth]{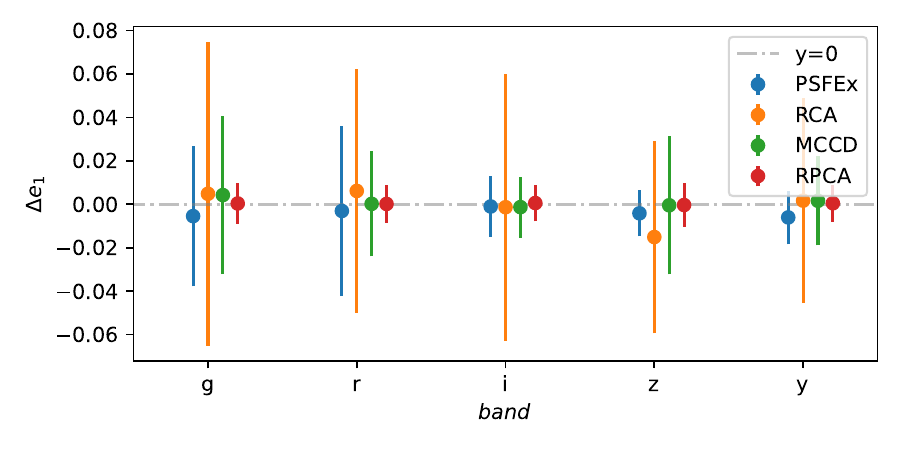}\\
\includegraphics[width=0.45\textwidth]{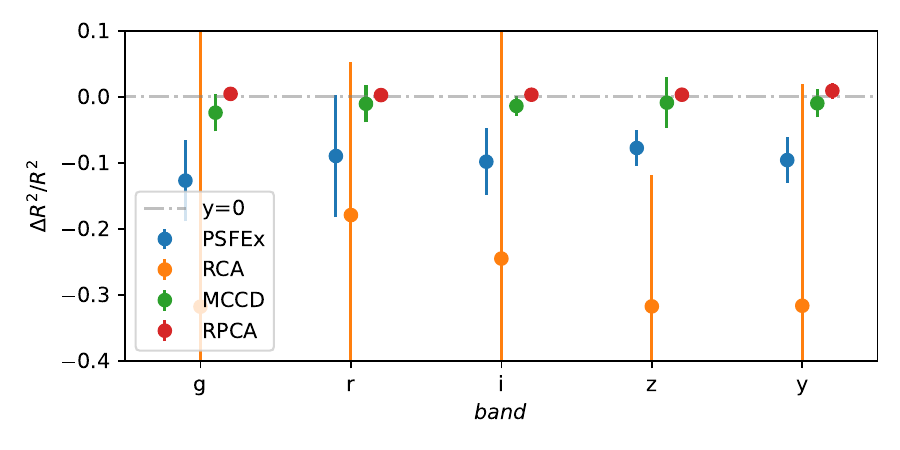}
\caption{Statistical shape residuals of  PSFEx (blue), RCA (orange), MCCD (green), and RPCA (red) are shown for the CSST MSC simulated image dataset. The panels from top to bottom represent the shape residuals: $\Delta e_1$, $\Delta e_2$, and $\Delta R^2/R^2$, respectively. The x-axis and y-axis represent the results from different bands and the statistical values of the shape parameters, respectively. The error bars indicate the $1\sigma$ width of the distributions.}
\label{mscshapes}
\end{figure}

\subsubsection{Metric analysis}
\label{sec:metric}
In order to investigate the PSF reconstruction performance of RPCA in more detail, we examine the shape parameters: size $R$ and ellipticity $e$ of the {interpolated} PSFs. These parameters are core indicators of PSF accuracy in weak lensing studies \citep{2018ARA&A..56..393M}. The ellipticity $e$ and size $R$ are typically defined based on the second-order central brightness moments $Q_{ij}$ of the PSF profile \citep{1998ApJ...504..636H}:
\begin{eqnarray}
e_1 = \frac{Q_{11} - Q_{22}}{Q_{11} + Q_{22}}, \quad e_2 = \frac{2Q_{12}}{Q_{11} + Q_{22}}, \quad \text{and} \quad R^2 = Q_{11} + Q_{22}. \nonumber \\
\label{shape}
\end{eqnarray}

Furthermore, we employ a circular Gaussian weight function to suppress the noise effect in the reconstructed PSF. The FWHM of the weight function is approximately 4.7 pixels at the instrument resolution and 9.4 pixels at the oversampled resolution, roughly twice the FWHM of the data. The differences in size and ellipticity between the input models and the {interpolated} PSFs, $\Delta e_i, i = [1, 2]$, and $\Delta R^2/R^2$, are used to quantify the precision of the reconstructed PSF models. Here, $i$ represents the two ellipticity components, and $R^2$ is measured from the fiducial {noiseless} PSF images.

Figure~\ref{ushapes} shows the statistical shape residual comparisons between the four methods based on the `$u$' band dataset with different SNRs. Additionally, we include a dataset with SNRs randomly distributed between 100 and 300 to evaluate the performance of our method under more realistic conditions. All methods can reconstruct the $e_1$ component without bias based on the instrument resolution dataset. The performance of RPCA is affected visibly by the data quality; in contrast, the errors of $\Delta e_1$ from MCCD and RCA are more stable as the SNR varies than those of RPCA. PSFEx appears to reconstruct the PSF $\Delta e_2$ and $R^2$ with slight biases, as shown in the lower-left panels, while RCA exhibits the largest error bars among the four methods. RPCA also provides the most accurate size reconstruction among the four methods.

For the comparison based on the oversampled datasets, all methods show larger errors in shape parameter reconstructions than in the instrument resolution case. However, RPCA still demonstrates the most stable results with the smallest bias, as shown in the right panels. The differences between RPCA, MCCD, RCA and PSFEx become more pronounced in the $\Delta e_2$ and $R^2$ reconstructions, as illustrated in the bottom two panels.

{Additionally, to assess the performance of the methods at the pixel level, we examine the mean residual images. Figure~\ref{uRMSE} shows the mean residual images for the `$u$' band dataset, with SNR varying randomly in the range 100-300, for PSFEx, RCA, MCCD, and RPCA under both instrument and oversampled conditions. RPCA exhibits the weakest residual structures among the methods. Furthermore, we use the pixel root-mean-square error (RMSE) \citep{2021A&A...646A..27L},  which is computed over all test images and then averaged, to quantify the performance of each method. The RMSE values are indicated in Figure~\ref{uRMSE}, and RPCA shows the lowest value among these methods. The performance on other datasets is similar to that in Figure~\ref{uRMSE}, and therefore we do not display them here.}

Figure~\ref{zshapes} shows the same comparisons as in Figure~\ref{ushapes}, but for the `$z$' band. All methods exhibit similar results to those in Figure~\ref{ushapes}. In the oversampled situation, RPCA shows larger error bars than MCCD and RCA based on the SNR=10 dataset. Nevertheless, RPCA still represents the most stable results among the four methods, with significantly smaller overall errors.  

{Figure~\ref{zRMSE} shows the mean residual images for the `$z$' band dataset, with SNR varying randomly in the range 100-300, for PSFEx, RCA, MCCD, and RPCA under both instrument and oversampled conditions. RPCA also exhibits the weakest residual structures among the methods. The RMSE values are indicated in Figure~\ref{zRMSE}, and RPCA again shows the lowest value among these methods. The performance on other datasets is similar to that in Figure~\ref{zRMSE}, and therefore we do not display them here.}

{From the results above, RPCA shows the lowest bias in shape and pixel value reconstruction based on our mock dataset. The size bias and error for RPCA can be controlled down to parts per thousand, especially for the `$z$' band, whereas for the other methods, the bias can reach the level of several percent. Therefore, RPCA reduces the bias of $\Delta R^2/R^2$ by approximately one order of magnitude compared to MCCD, RCA, and PSFEx. However, the size bias and error for the `$u$' band are larger for all methods due to the poorer sampling conditions.}

\subsection{Test on CSST MSC simulation data}

In order to investigate more realistic effects, we use the simulated images from the CSST Main Survey Camera to evaluate the performance of our method on a more realistic dataset. The image simulations take into account various potential effects, including the spectral energy distribution (SED), magnitude, and positions of different sources, sky background, dark current, readout noise, bad pixels of the CCD detector, filters, cosmic ray contamination, and other factors. In addition, the brighter-fatter effect \citep{2014JInst...9C3048A} added in the simulation is very weak currently; therefore, we neglect this effect in this work.

We utilize the under-development CSST data processing pipeline to subtract the background noise, mask cosmic rays and bad pixels, and extract sources.
Furthermore, the data from the two short-wavelength bands, `nuv` and `u`, are not included in this section due to the limited number of stars in these bands. Owing to the small number of stars in the images, only star images with SNRs $\geq$ 40 are extracted for PSF reconstruction tests, resulting in an average of 60 stars identified per CCD chip. In the MSC image simulation, we only have the noisy data with {various effects} added, hence we use the extracted star images as the reference PSFs. Figure \ref{mscstars} shows some extracted star images from the CSST MSC Simulation image.

The shape residuals from the four methods are shown in Figure~\ref{mscshapes}. The reconstructed shape precisions are consistent with the results presented in Figures~\ref{ushapes} and \ref{zshapes}, {and MCCD shows much  better size reconstructions at observed positions for instrument resolution on the MSC simulation data}. The errors of $\Delta R^2/R^2$ for RPCA are reduced by a factor of approximately 10 compared to PSFEx and RCA. Therefore, RPCA demonstrates robust stability under real data quality conditions and shows promising potential for future applications in CSST PSF reconstructions.

\section{Conclusions}\label{sec:sums}
High-resolution images from space telescopes like CSST will provide new insights into our universe. However, the potential undersampled images from space instruments make the PSF modeling process challenging, and scientific research based on PSF models—such as weak lensing and galaxy morphology analysis—becomes more difficult. To construct reliable PSFs for poorly resolved images, this paper focuses primarily on the first step of the PSF modeling process.

We transform the subpixel shifting and downsampling process into matrix operations, providing a direct way to solve for the oversampled PSF. To avoid potential bias caused by inappropriate PSF variation models across the field during the reconstruction process, we relax the requirement for position correlation in our PSF reconstruction. The methodology of EMPCA is adopted in our approach due to its flexible handling of noisy and missing data, which is well-suited for practical observational image data.

We evaluate the performance of our method using high-fidelity simulated CSST PSF data and MSC simulation images. The pixel residuals of RPCA are reduced to parts per thousand for both instrument and oversampled datasets, whereas those of PSFEx and RCA  could exceed 0.01. Furthermore, we assess the precision of the reconstructed PSFs using shape parameters: $\Delta e_1$, $\Delta e_2$, and $\Delta R^2/R^2$. RPCA demonstrates the most unbiased mean shape deviations with errors around several percent for the instrument dataset, while MCCD, RCA, and PSFEx exhibit more significant biases in $\Delta R^2/R^2$ along with larger errors. The results based on the oversampled dataset indicate that the reconstructed PSF shape precision of RPCA remains within several percent, suggesting its potential for image deconvolution \citep{2024RAA....24d5009W,2025ApJS..276...36W}.

Overall, the results demonstrate that RPCA achieves the highest precision in PSF shape parameters compared to MCCD, RCA, and PSFEx. In particular, for $\Delta R^2/R^2$, RPCA could reduce the errors by approximately an order of magnitude compared to MCCD, RCA, and PSFEx, highlighting its promising prospects for future applications in CSST. {For fairness, we note that MCCD, originally designed for Multi-CCD Point Spread Function modeling, is applied here in a single-CCD scenario due to the current camera configuration.} Additionally, the anti-aliasing capability of RPCA provides a new avenue for high-resolution reconstruction based on low-resolution spectral data \citep{2006ApJS..163..110S,2024arXiv240311011H}.

However, as mentioned earlier, the spatial variations of PSFs across the field need to be considered in the interpolation scheme \citep{2013A&A...549A...1G}, and the simple polynomial used in this work may not be the best choice for CSST PSF interpolation. Modeling these variations is also a challenging problem, related not only to the instrument-induced variations of the PSFs themselves but also to the stellar number density \citep{2008A&A...484...67P,2009A&A...500..647P}. In the future, we will investigate these effects and explore various interpolation methods to develop a reliable interpolation scheme combined with RPCA.

Furthermore, the color dependence of PSFs is not considered in this paper. A more suitable approach to address this issue may involve generating PSF models that incorporate SED information \citep{2015ApJ...807..182M,2023InvPr..39c5008L}, followed by applying corrections to the model using RPCA, similar to the work presented in \citet{2024AJ....167...58N}. However, \citet{2023RAA....23g5021L} reports that the color dependence for CSST is not significant. Overall, RPCA provides a simple and effective way to reconstruct reliable PSF models using only star images.

\section*{Data availability}
The code and associated data can be found at github: \url{https://github.com/linngithub0/Resolved-PCA/tree/main} and Zenodo: \url{https://doi.org/10.5281/zenodo.23208591}.

\begin{acknowledgements}
      The authors thank the referees, whose comments help us to significantly improve the presentation and understanding of the method. This work is supported by National Key R$\&$D Program of China No. 2022YFF0503403 and the Ministry of Science and Technology of China (grant Nos. 2020SKA0110100). H.Y.S. acknowledges the support from NSFC of China under grant 11973070, Key Research Program of Frontier Sciences, CAS, grant No. ZDBS-LY-7013 and Program of Shanghai Academic/Technology Research Leader. L.N. acknowledges the support from NSFC of China under grant 12503009. We acknowledge the support from the science research grants from the China Manned Space Project with No. CMS-CSST-2021-A01, CMS-CSST-2021-A04. We also acknowledge the helpful discussions with Dezi Liu and Charling Tao.\\

    Software: astropy \citep{2013A&A...558A..33A,2018AJ....156..123A,2022ApJ...935..167A},  
          Source Extractor \citep{1996A&AS..117..393B},
          PyTorch \citep{NEURIPS2019_bdbca288},
          PSFEx \citep{2011ASPC..442..435B},
          GalSim \citep{2015A&C....10..121R},
          SciPy \citep{2020NatMe..17..261V},
          NumPy \citep{harris2020array},
          Matplotlib \citep{2021zndo....592536T}
\end{acknowledgements}

   \bibliographystyle{aa}
   \bibliography{RPCA}
\begin{appendix}
\section{Matrix operation for convolution and downsampling}
\label{sec:appa}
{In image processing, we usually treat the image as a 2-dimensional matrix:
\[
\begin{matrix}
\boldsymbol{y} =
\begin{bmatrix}
\cdots & \cdots & \cdots & \cdots & \cdots\\
\cdots & y(u+1,v-1) & y(u+1,v) & y(u+1,v+1) &\cdots\\
\cdots & y(u,v-1) & y(u,v) & y(u,v+1) &\cdots\\
\cdots & y(u-1,v-1) & y(u-1,v) & y(u-1,v+1) &\cdots\\
\cdots & \cdots & \cdots & \cdots & \cdots
\end{bmatrix}
\end{matrix}
\]
where $\boldsymbol{y}$ is the image to be processed. The convolution kernel $\boldsymbol{K}$ can be expressed similarly:
\[
\begin{matrix}
\boldsymbol{K} =
\begin{bmatrix}
\cdots & \cdots & \cdots & \cdots & \cdots\\
\cdots & K(1,-1) & K(1,0) & K(1,1) &\cdots\\
\cdots & K(0,-1) & K(0,0) & K(0,1) &\cdots\\
\cdots & K(1,-1) & K(-1,0) & K(-1,1) &\cdots\\
\cdots & \cdots & \cdots & \cdots & \cdots
\end{bmatrix}
\end{matrix}
\]
Generally, $(0,0)$ represents the center of the kernel, which should be aligned with the pixel $(u^\prime,v^\prime)$ in Equation~\ref{conv}. Equation~\ref{conv} is essentially an inner product between a 2-dimensional matrix $\boldsymbol{S^{\prime}}(u,v)$ and $\boldsymbol{y}$:
\begin{eqnarray}
y^{\prime}(u,v) = \boldsymbol{y} \cdot {S^{\prime}}(u,v) = {S^{\prime}}^T(u,v) \times \boldsymbol{y},
\end{eqnarray}
where $\boldsymbol{S^{\prime}}(u,v)$ is the centrosymmetric matrix of $\boldsymbol{K}(u,v)$, and $\cdot$ denotes the Frobenius inner product.
If we treat $\boldsymbol{y}$ and $\boldsymbol{S^{\prime}}$ as 1-dimensional vectors, then $y^{\prime}(u,v)$ becomes the standard vector inner product of $\boldsymbol{y}$ and the transpose of ${S^{\prime}}$: ${S^{\prime}}^T$. Figure~\ref{fig:conv} illustrates this concept, and $\boldsymbol{\rm S}$, which is formed by stacking all ${S^{\prime}}(u,v)$ for every pixel, represents the interpolation matrix in Equation~\ref{conv}.}

{Generally, Lanczos or cubic functions are widely applied in astronomical image interpolation \citep{2011ASPC..442..435B,2021MNRAS.501.1282J} due to potential undersampling effects. \citet{2014PASP..126..287B} shows that the sinc function is a lossless interpolation function for band-limited signals, and for well-resolved PSF images, we expect the PSF to be well sampled. Therefore, we use the sinc function to construct the convolution kernel for subpixel shifts. Additionally, \citet{1999SPIE.3661..728C} shows that the sinc interpolation kernel constructed using FFT can achieve high precision, providing an efficient way to generate the kernel. Hence, we also use FFT to generate the kernel for subpixel shifts, where the real-space kernel is obtained by applying the inverse FFT on the kernel in Fourier space.}

{Similarly, the down-sampling process can be represented as a matrix inner product. Figure~\ref{fig:downsam} illustrates this matrix formulation. For non-integer down-sampling factors, the weights of subpixel areas must be included in the down-sampling matrix, which adds complexity; we plan to consider this effect in future work. Furthermore, the non-linear response within individual pixels is not taken into account in this work, but it could be incorporated by modifying the values in the $\boldsymbol{D}$ matrix, which we also plan to test in the future.}

{Finally, both subpixel shifting and down-sampling can be formulated as matrix operations, as illustrated in Figure~\ref{fig:combined}.}

\begin{figure}
    \centering
    \includegraphics[width=.8\linewidth]{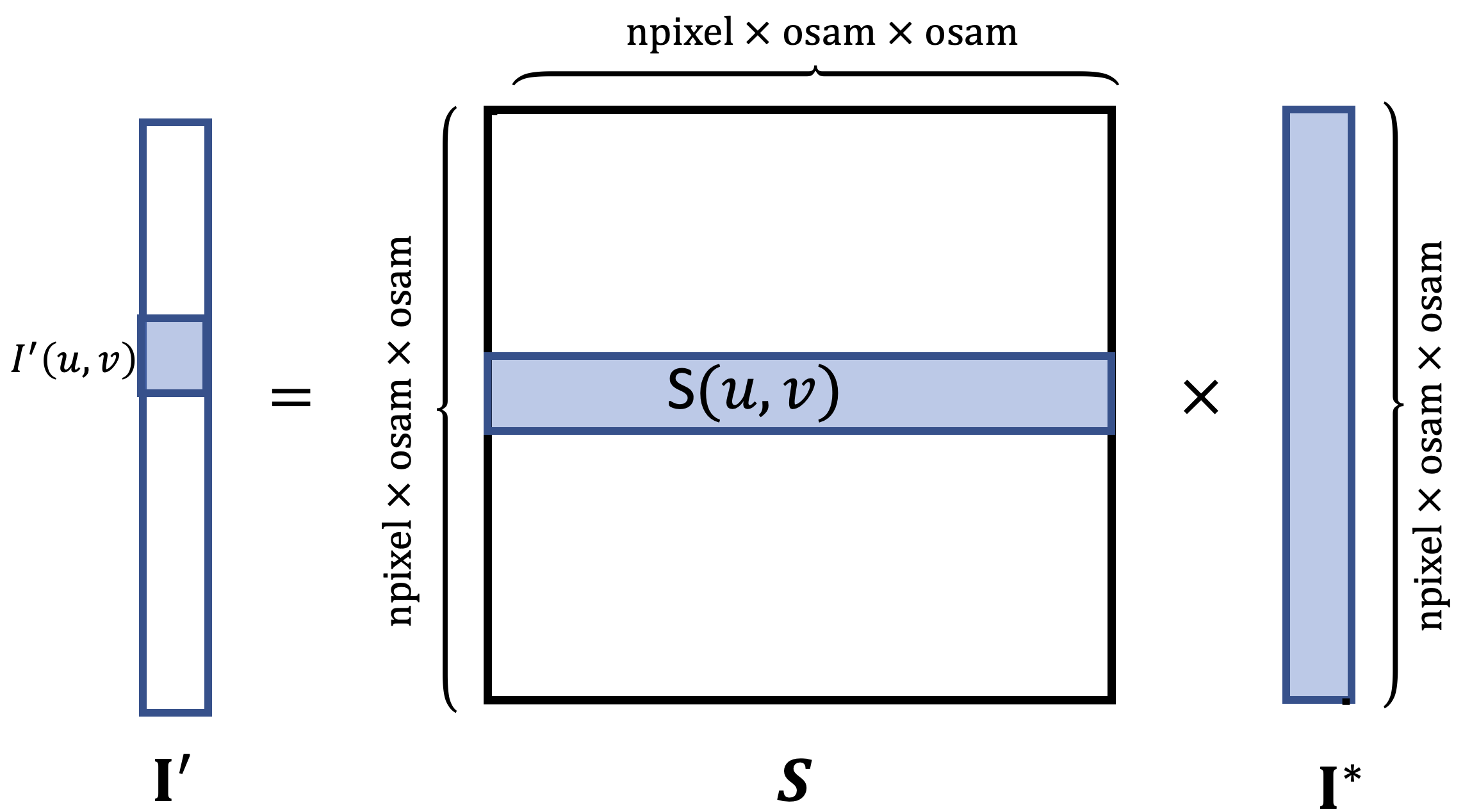}
    \caption{Illustration of the matrix formulation for convolution. $\boldsymbol{\rm S}$ is the convolution matrix in Eq.~(\ref{conv}), and $S(u, v)$ represents the flattened convolution kernel for $I^{\prime}(u,v)$. $\boldsymbol{\rm I^{\ast}}$ is the original image vector in Eq.~(\ref{conv}). osam denotes the oversampling factor.}
    \label{fig:conv}
\end{figure}

\begin{figure}
    \centering
    \includegraphics[width=.8\linewidth]{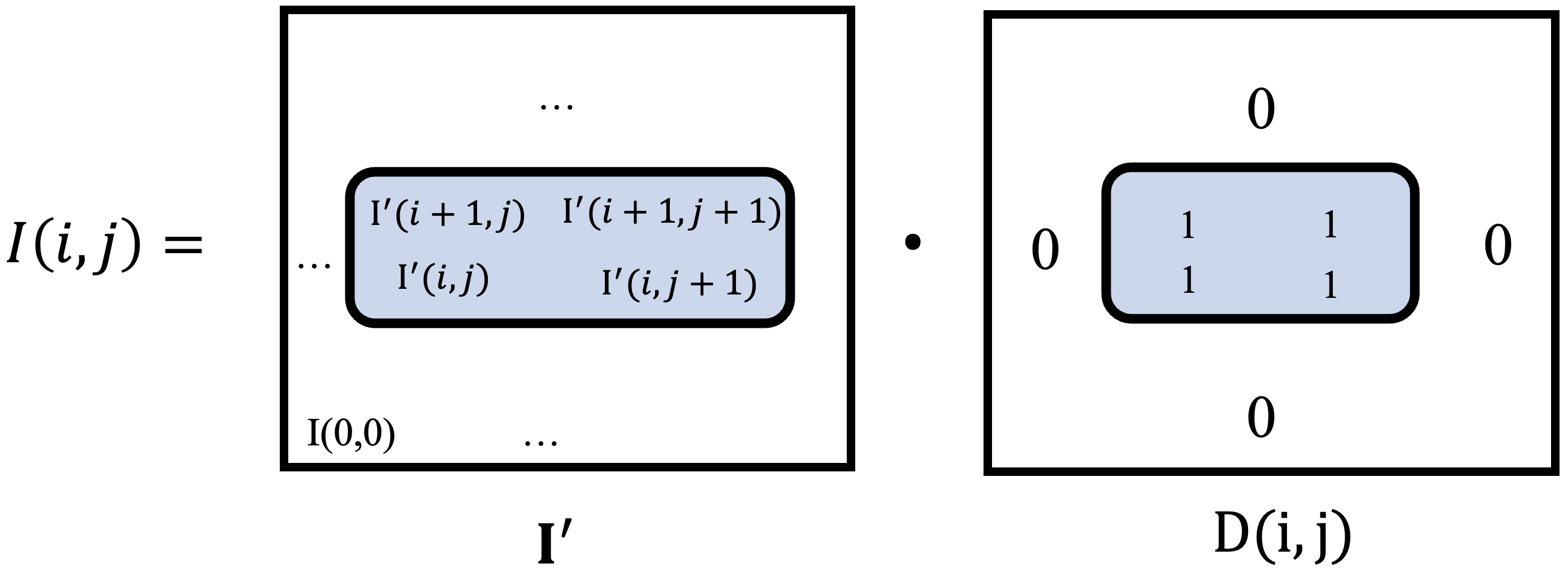}
    \includegraphics[width=.8\linewidth]{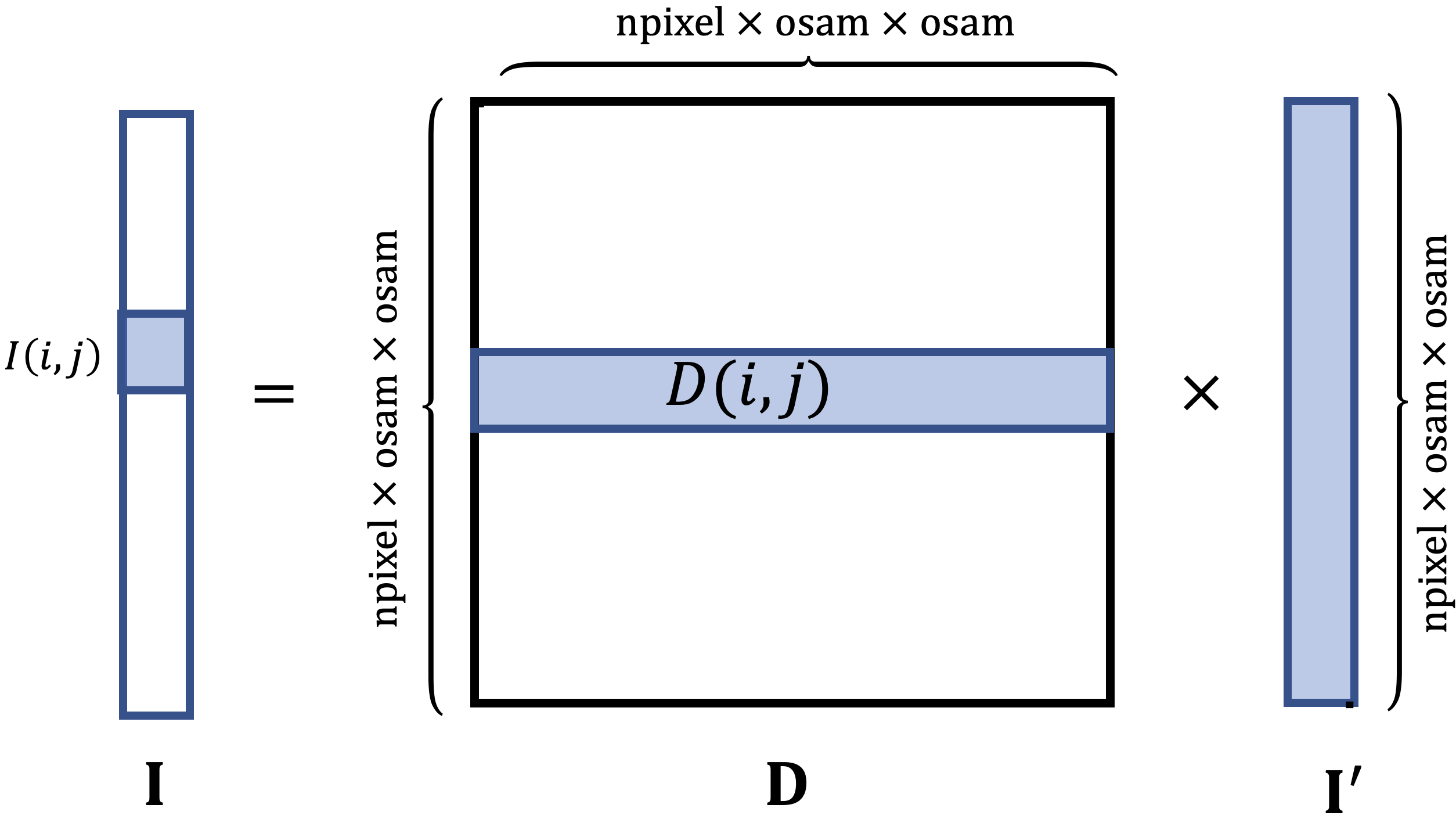}
    \caption{Illustration of the matrix formulation for down-sampling with $osam=2$.
Upper panel: The pixel value of $I(i,j)$ in the down-sampled image is given by: $I(i,j)=I^{\prime}(i,j)+I^{\prime}(i,j+1)+I^{\prime}(i+1,j)+I^{\prime}(i+1,j+1)$, which can also be regarded as the inner product of the matrix $\boldsymbol{I^{\prime}}$ and the matrix $D(i,j)$, where $D(i,j)$ is the specific matrix corresponding to $I(i,j)$.
Lower panel: The pixel value of $I(i,j)$ computed via matrix operations, where the row $D(i,j)$ and $\boldsymbol{\rm I^{\prime}}$ represent the flattened vectors of the matrices.} 
    \label{fig:downsam}
\end{figure}

\begin{figure}
    \centering
    \includegraphics[width=1.\linewidth]{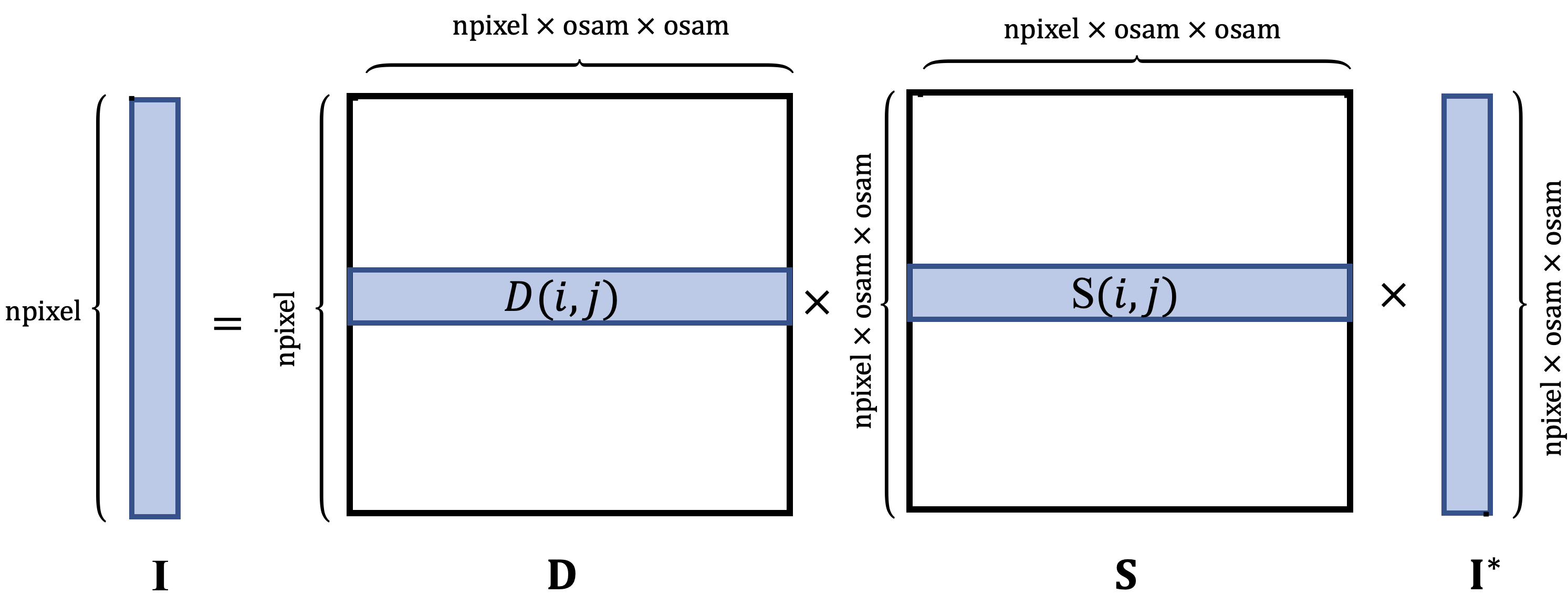}
    \caption{Illustration of the matrix process for convolution and down-sampling.}
    \label{fig:combined}
\end{figure}

\clearpage
\section{Parameters chosen for RCA and MCCD}
\label{sec:para-check}

{Several hyperparameters need to be configured for RCA and MCCD. The main parameters affecting the performance of the constructed PSF models are: the number of local components $n\_comp\_loc$, the regularization parameter $ksig$, and the local hybrid degree $d\_hyb\_loc$ (only for MCCD). To determine appropriate parameter settings for our tests, we ran multiple experiments with different combinations of these parameters for both methods. Table~\ref{tab:para-check} lists the parameter space explored for the simulated PSF data and the MSC simulation data, respectively.
To evaluate the performance of each parameter configuration, we use the sum of shape biases, $\langle \Delta e \rangle + \langle \Delta R^2 / R^2 \rangle$, as a metric. Figures~\ref{fig:RCA-para} and \ref{fig:MCCD-para} show the performance of different parameter settings for RCA and MCCD on the simulated PSF data and MSC simulation data, respectively. Based on these results, we adopt $n\_comp\_loc=10, ksig=1$ for RCA and $n\_comp\_loc=5, d\_hyb\_loc=3, ksig=3$ for MCCD on the simulated PSF dataset. For the MSC simulation data, we adopt $n\_comp\_loc=5, ksig=1$ for RCA and $n\_comp\_loc=15, d\_hyb\_loc=2, ksig=1$ for MCCD.}
\begin{table}
\centering
\caption{Parameter space explored for RCA and MCCD with an oversampling factor of 2.}
\label{tab:para-check}
\begin{tabular}{lcr}
\hline
\hline
Parameter & RCA               & MCCD    \\
\hline
$n\_comp\_loc$                 & 5, 10, 15, 20   & 5, 10, 15     \\
$ksig$                                 & 0.0, 1.0, 3.0    & 0.0, 1.0, 3.0  \\
$d\_hyb\_loc$                   & $\cdots$         & 2, 3, 4          \\
\hline
\hline
\end{tabular}
\end{table}

\begin{figure}
    \centering
    \includegraphics[width=.5\linewidth]{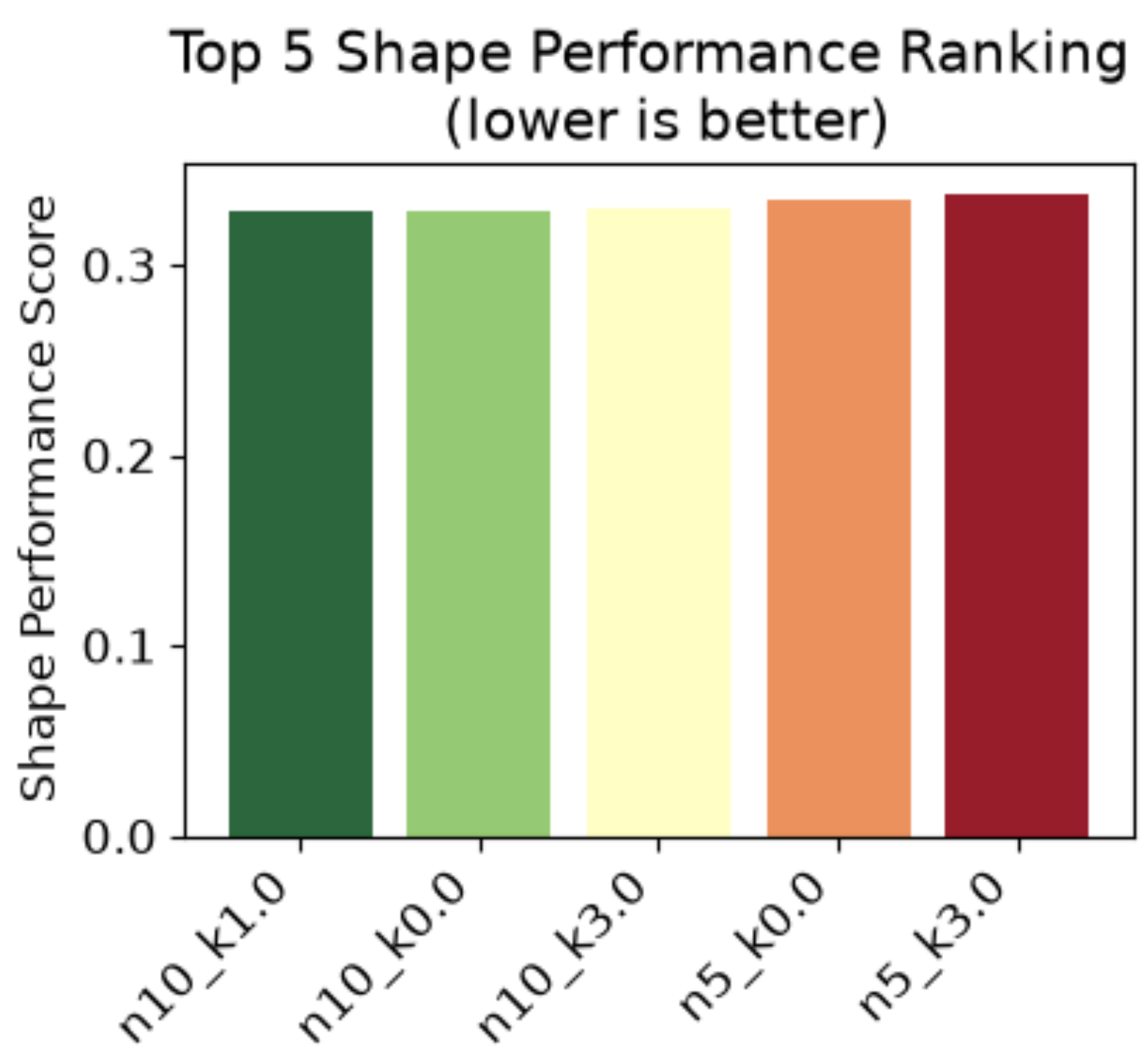}
    \includegraphics[width=.5\linewidth]{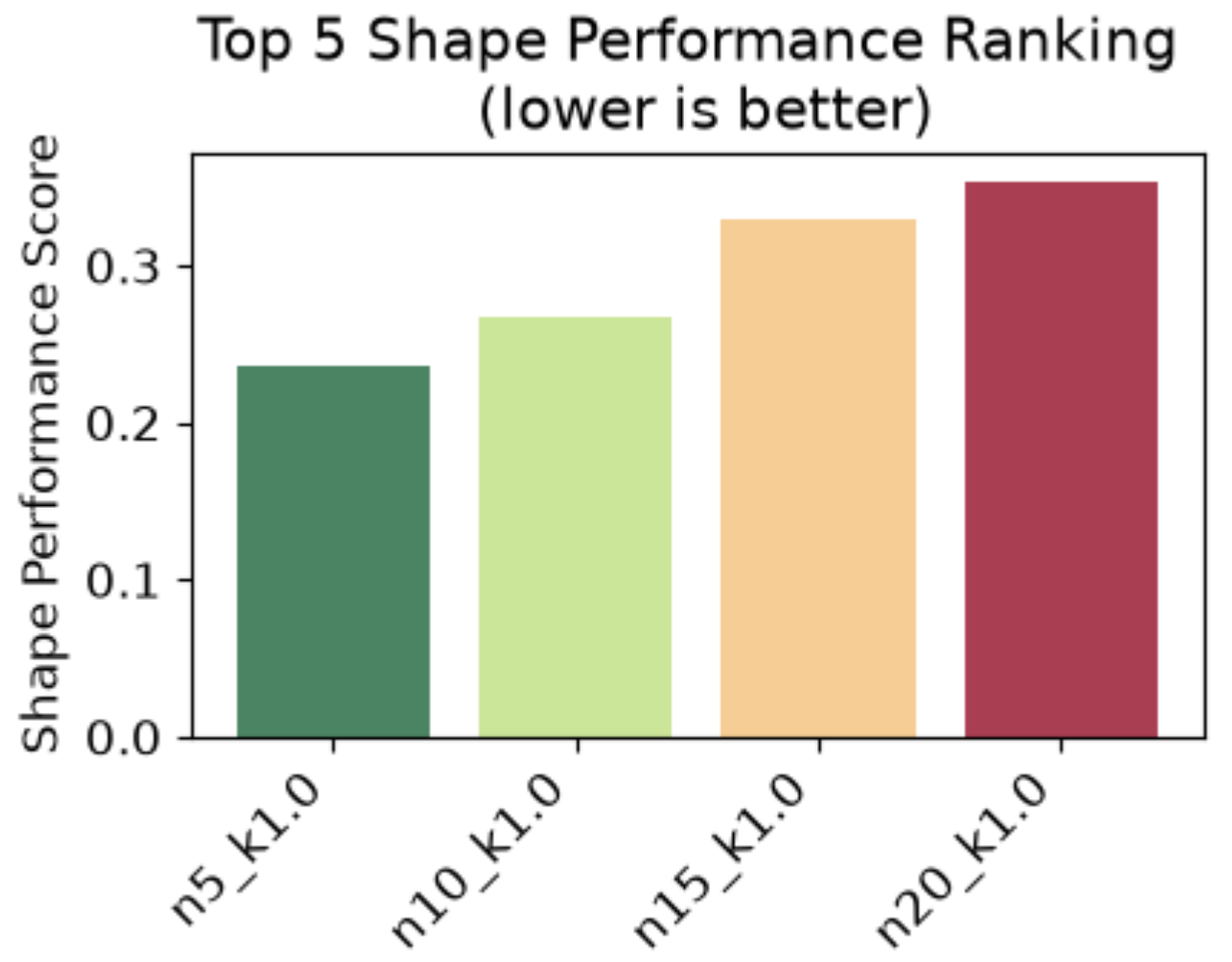}
    \caption{Performance of RCA with different parameter settings on the simulated PSF data and MSC simulation data. On the x-axis, the numbers following n and k indicate the number of local components and the initial regularization parameter, respectively.
Upper panel: Shape performance for different parameter settings on the simulated PSF data.
Lower panel: Shape performance for different parameter settings on the MSC simulation data.}
    \label{fig:RCA-para}
\end{figure}

\begin{figure}
    \centering
    \includegraphics[width=.5\linewidth]{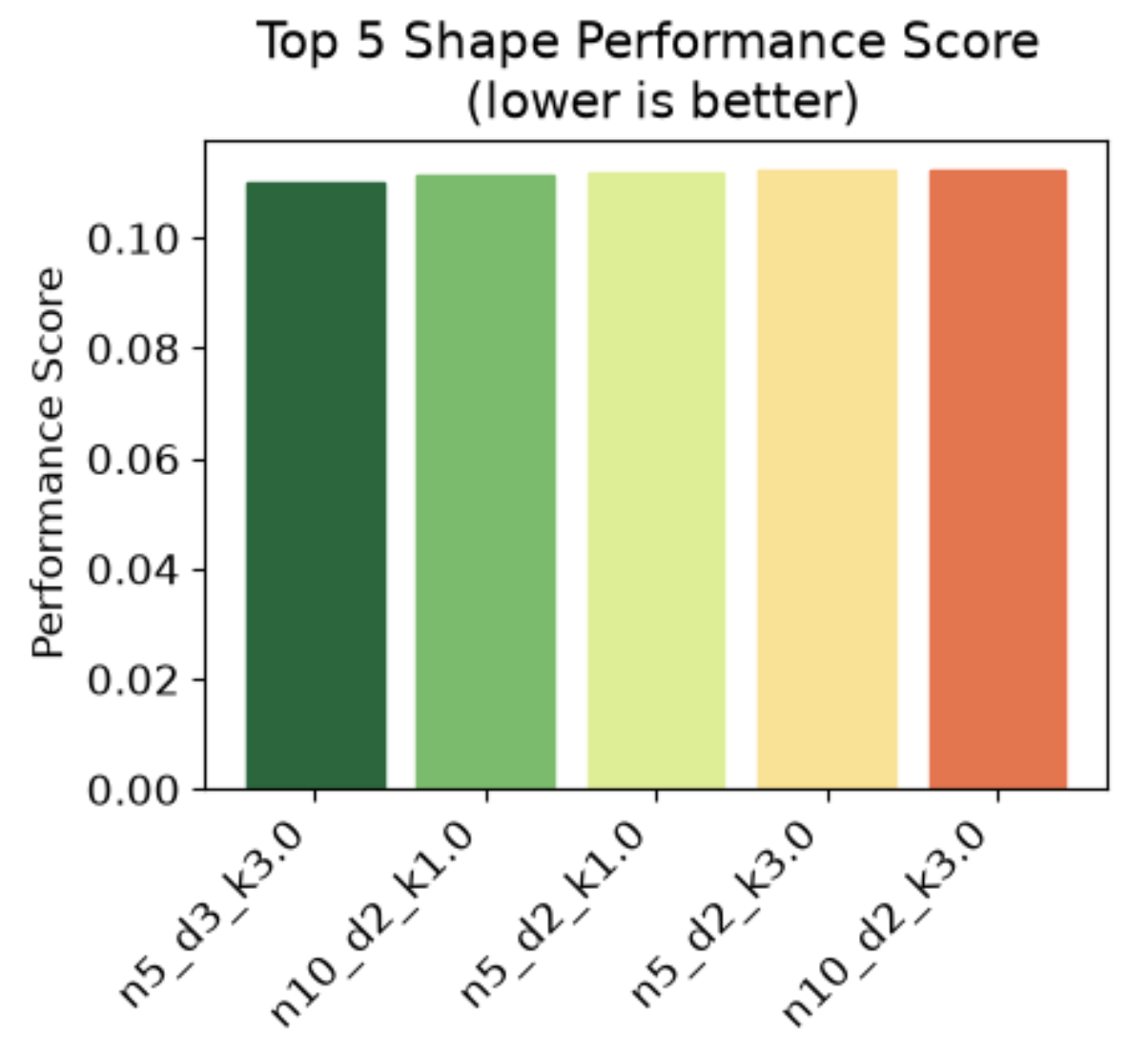}
    \includegraphics[width=.5\linewidth]{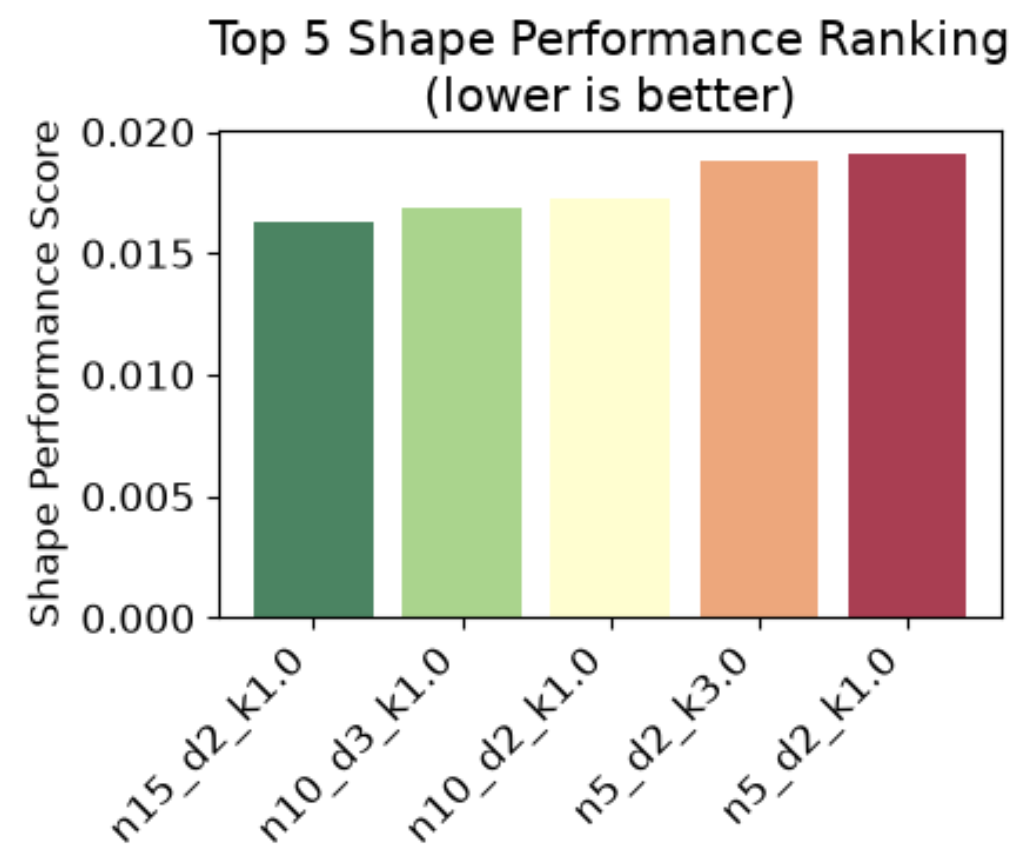}
    \caption{Performance of MCCD with different parameter settings on the simulated PSF data and MSC simulation data. On the x-axis, the numbers following n, d, and k indicate the number of local components, the local hybrid degree, and the initial regularization parameter, respectively.
Upper panel: Shape performance for different parameter settings on the simulated PSF data.
Lower panel: Shape performance for different parameter settings on the MSC simulation data.}
    \label{fig:MCCD-para}
\end{figure}

\onecolumn

\section{Shape residuals in the z band}

\begin{figure*}[h!]
\includegraphics[width=0.5\textwidth]{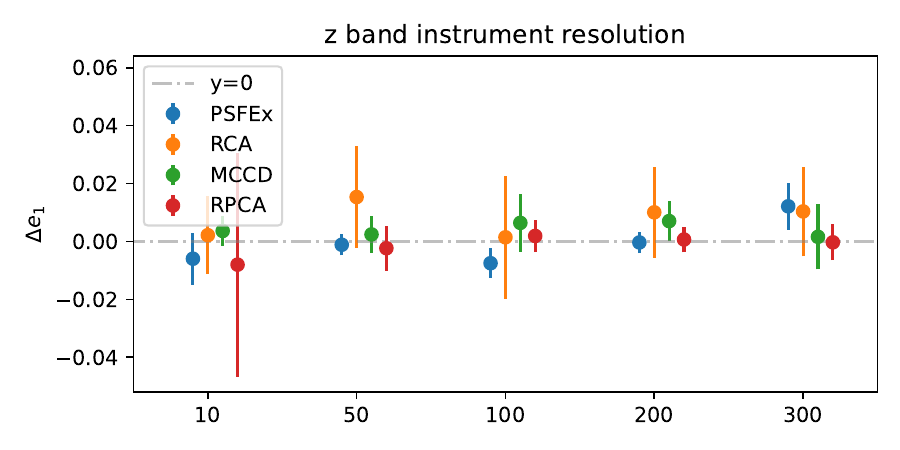}
\includegraphics[width=0.5\textwidth]{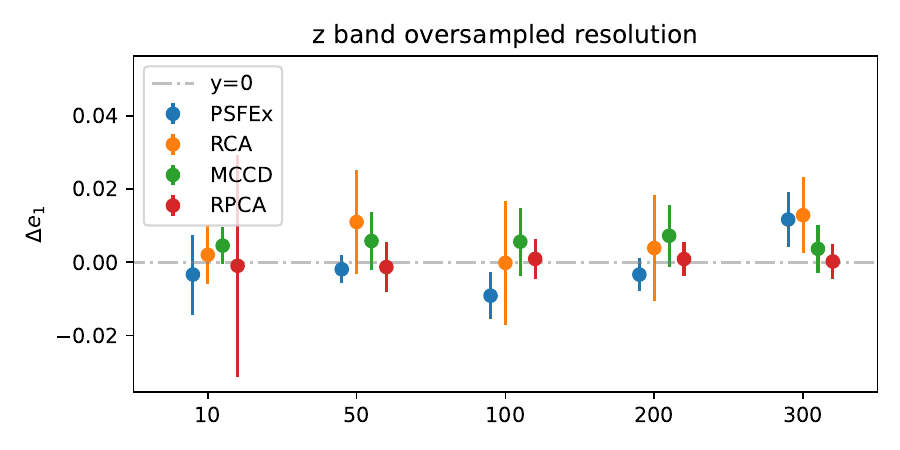}
\includegraphics[width=0.5\textwidth]{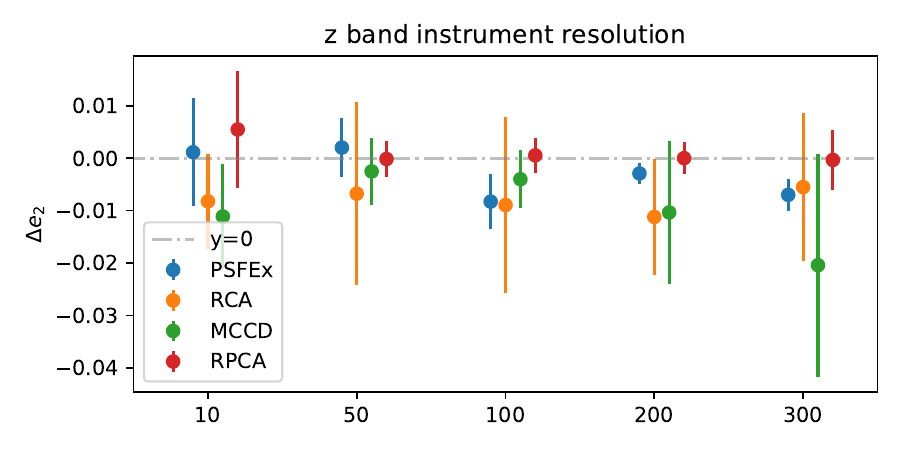}
\includegraphics[width=0.5\textwidth]{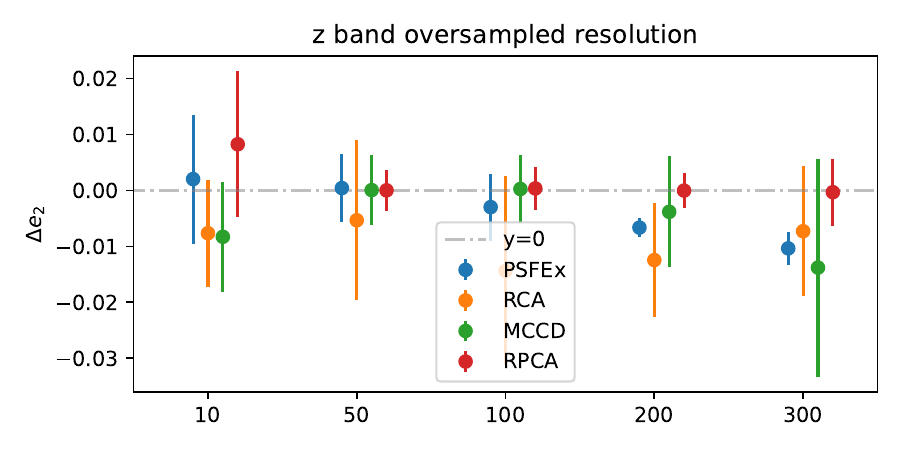}
\includegraphics[width=0.5\textwidth]{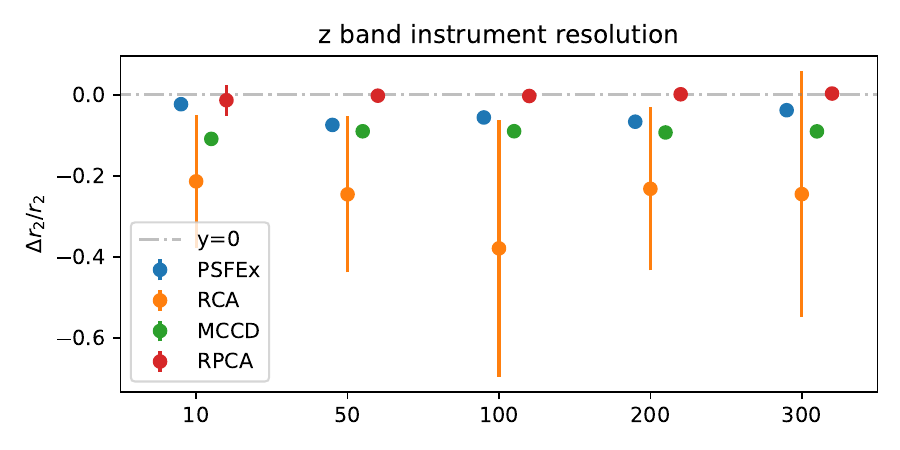}
\includegraphics[width=0.5\textwidth]{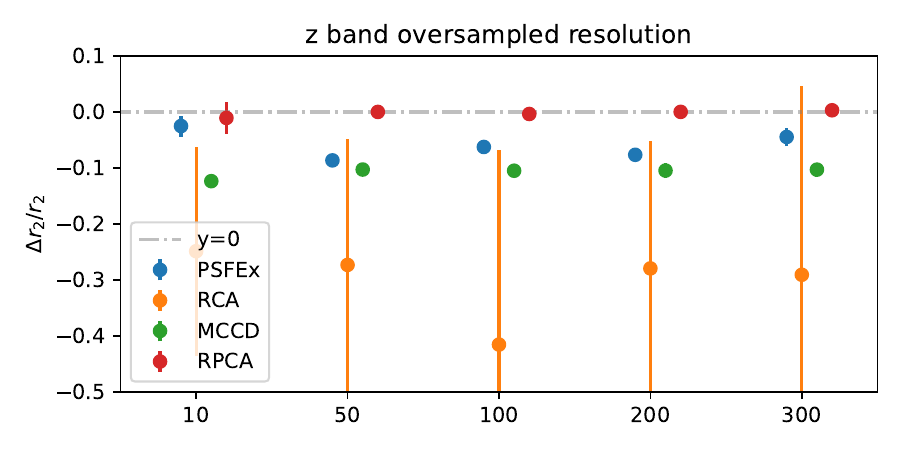}
\caption{The same as in Fig.~\ref{ushapes} but in the `z' band.}
\label{zshapes}
\end{figure*}

\FloatBarrier

\twocolumn

\end{appendix}

\end{document}